\documentclass[journal]{IEEEtran}

\usepackage{bm}
\usepackage{url}
\usepackage{color}
\usepackage{dsfont}
\usepackage{subfig}
\usepackage{wrapfig}
\usepackage{multirow}
\usepackage{booktabs} 
\usepackage{graphicx}
\usepackage{makecell}
\usepackage{enumitem}
\usepackage{algorithm}
\usepackage{algorithmic}
\usepackage[dvipsnames]{xcolor}
\usepackage{amsmath,amssymb,amsfonts}
\usepackage[font=footnotesize, skip=4pt]{caption}

\usepackage{tikz}
\usetikzlibrary{calc}
\usetikzlibrary{bayesnet}
\usetikzlibrary{positioning}
\usetikzlibrary{datavisualization}
\usetikzlibrary{decorations.pathreplacing,angles,quotes}
\usetikzlibrary{shapes,arrows,snakes}
\tikzstyle{status} = [rectangle, draw=black, text centered, anchor=north, text=black, minimum width=8em, minimum height=3em, node distance=6ex and 7em]
\tikzstyle{line} = [draw,thick,-latex]
\tikzstyle{transition} = [font=\small]
\tikzstyle{mybox} = [thick,
    rectangle, rounded corners, inner sep=2pt, inner ysep=-2pt]
\tikzstyle{fancytitle} =[text=white]
\tikzstyle{latent} = [circle,fill=white,draw=black,inner sep=1pt,
minimum size=12pt, font=\fontsize{7}{7}\selectfont, node distance=0.8]

\usepackage{fancyvrb}
\usepackage{framed, color}
\definecolor{shadecolor}{rgb}{1, 0.8, 0.3}
\definecolor{myblue}{rgb}{1, .8, 0.3}
\definecolor{mygreen}{rgb}{0.14, 0.5, 0.14}
\definecolor{myyellow}{rgb}{0.83, 0.83, 0}
\definecolor{mygrey}{rgb}{0.5, 0.5, 0.5}
\definecolor{myorange}{rgb}{1, 0.28, 0}
\definecolor{mytransp}{rgb}{0.8, 0.8 ,0.8}

\makeatletter
\let\NAT@parse\undefined
\makeatother
\usepackage[colorlinks,linkcolor=red]{hyperref}
\hypersetup{
  colorlinks=true,
  citecolor=blue}

\usepackage{amsthm}

\newtheorem{assumption}{Assumption}

\IEEEoverridecommandlockouts

\newcommand{\compresslist}{ 
\setlength{\itemsep}{1pt}
\setlength{\parskip}{0pt}
\setlength{\parsep}{0pt}
}

\makeatletter
\def\namedlabel#1#2{\begingroup
   \def\@currentlabel{#2}%
   \label{#1}\endgroup
}

\makeatother

\begin{document}
\title{Estimating Heterogeneous Treatment Effects in Residential Demand Response}
\author{Datong~P.~Zhou$^\ast$, Maximilian~Balandat$^\dagger$, and~Claire~J.~Tomlin$^\dagger$,~\IEEEmembership{Fellow,~IEEE}
\thanks{$^\ast$Dept. of Mechanical Engineering, University of California, Berkeley (UCB). $^\dagger$Dept. of Electrical Engineering and Computer Sciences, UCB.\texttt{[datong.zhou, balandat, tomlin]@berkeley.edu}}
}

\markboth{Working Paper. Last updated: October 23, 2018}%
{Zhou \MakeLowercase{\textit{et al.}}: Estimating Heterogeneous Treatment Effects in Residential Demand Response}

\maketitle

\begin{abstract}
We evaluate the causal effect of hour-ahead price interventions on the reduction in residential electricity consumption using a data set from a large-scale experiment on 7,000 households in California. By estimating user-level counterfactuals using time-series prediction, we estimate an average treatment effect of $\approx$ 0.10 kWh (11\%) per intervention and household. Next, we leverage causal decision trees to detect treatment effect heterogeneity across users by incorporating census data. These decision trees depart from classification and regression trees, as we intend to estimate a causal effect between treated and control units rather than perform outcome regression. We compare the performance of causal decision trees with a simpler, yet more inaccurate $k$-means clustering approach that naively detects heterogeneity in the feature space, confirming the superiority of causal decision trees. Lastly, we comment on how our methods to detect heterogeneity can be used for targeting households to improve cost efficiency.
\end{abstract}


\IEEEpeerreviewmaketitle

%
%

\section{Introduction}
This paper studies the causal effect of incentivizing residential households to participate in Demand Response (DR) to temporarily reduce electricity consumption. DR has been promoted by the introduction of demand-side management programs (DSM) after the 1970s energy crisis \cite{Palensky:2011aa}, enabled by the integration of information and communications technology in the electric grid. The rationale behind DSM is to alleviate the inelasticity of energy supply due to the slowness of power plants' output adjustment, which causes small increases and decreases in demand to result in a price boom or bust, respectively. DSM attempts to protect utilities against such price risks by partially relaying them to end-users, increasing market efficiency \cite{Borenstein:2005aa}.

\begin{figure}[h]
\centering
\vspace*{-0.45cm}
\begin{tikzpicture}
\node [status, draw=none, align=left, inner sep=2pt, minimum size=1pt] (WSM) {\textbf{\footnotesize{Wholesale Market}} \includegraphics[width=0.3cm]{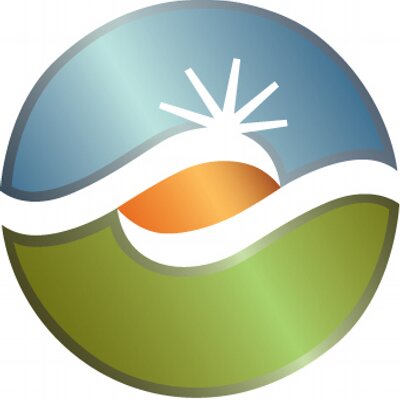}};

\node [status, draw=none, below=1.3em of WSM, xshift=-3cm, inner sep=2pt, minimum size=1pt] (DRP) {\textbf{\footnotesize{DR Provider}} \includegraphics[width=0.3cm]{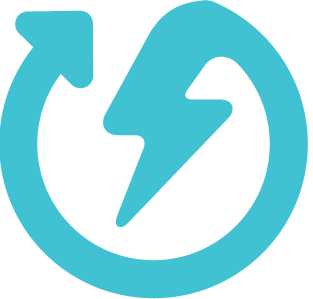}};

\node [status, draw=none, below=0.7em of WSM, xshift=3cm, inner sep=2pt, minimum size=1pt] (SC) {\textbf{\footnotesize{Scheduling Coordinator}} \includegraphics[width=0.5cm]{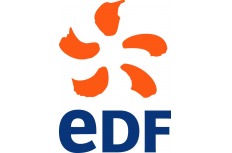}};

\node [status, draw=none, below=3.5em of WSM, xshift=3cm, inner sep=2pt, minimum size=1pt] (Ut) {\textbf{\footnotesize{Electric Utility}} \includegraphics[width=0.3cm]{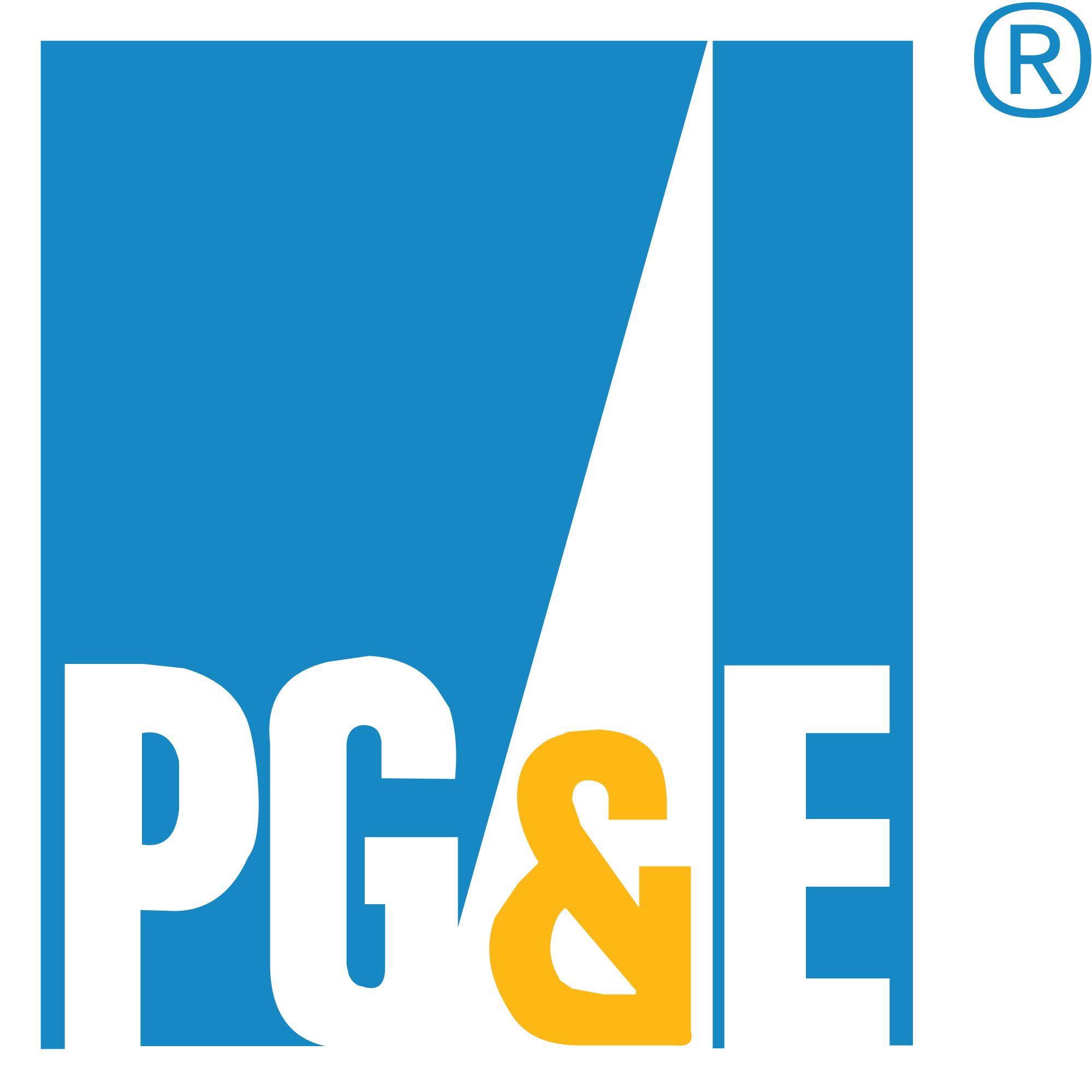}\includegraphics[width=0.4cm]{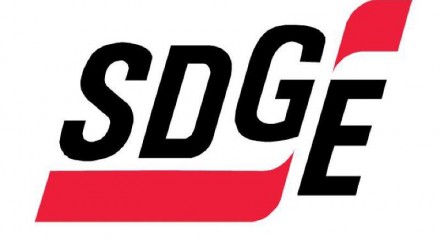}\includegraphics[width=0.3cm]{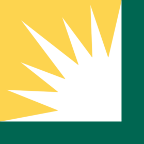}};

\node [status, draw=none, below=8.6em of WSM, inner sep=2pt, minimum size=1pt] (EU) {\textbf{\footnotesize{End-Use Customers}}};

\node[inner sep=0pt] (house1) at (-1.5,-3)
    {\includegraphics[width=0.4cm]{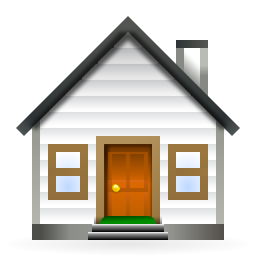}};
\node[inner sep=0pt] (house2) at (-1.1,-2.7)
    {\includegraphics[width=0.4cm]{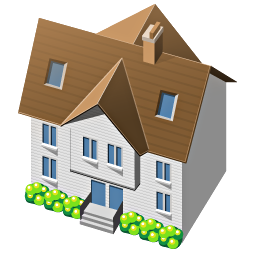}};
\node[inner sep=0pt] (house3) at (-0.7,-3)
    {\includegraphics[width=0.4cm]{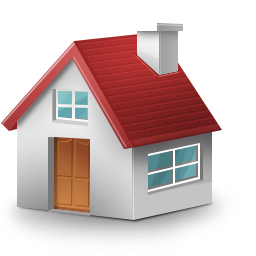}};
\node[inner sep=0pt] (house4) at (-0.3,-2.7)
    {\includegraphics[width=0.4cm]{Figures/House2.png}};
\node[inner sep=0pt] (house5) at (0.1,-3)
    {\includegraphics[width=0.4cm]{Figures/House3.png}};
\node[inner sep=0pt] (house6) at (0.5,-2.7)
    {\includegraphics[width=0.4cm]{Figures/House1.png}};
\node[inner sep=0pt] (house7) at (0.9,-3)
    {\includegraphics[width=0.4cm]{Figures/House1.png}};
\node[inner sep=0pt] (house8) at (1.3,-2.7)
    {\includegraphics[width=0.4cm]{Figures/House3.png}};
\node[inner sep=0pt] (house9) at (1.7,-3)
    {\includegraphics[width=0.4cm]{Figures/House2.png}};
        
\path [line, line width=0.15, <->, >=stealth] (WSM) -- (DRP);

\path [line, line width=0.15, <->, >=stealth] ([xshift=2em]DRP.south) -- (house1);
\path [line, line width=0.15, <->, >=stealth] ([xshift=2em]DRP.south) -- (house2);
\path [line, line width=0.15, <->, >=stealth] ([xshift=2em]DRP.south) -- (house3);
\path [line, line width=0.15, <->, >=stealth] ([xshift=2em]DRP.south) -- (house4);
\path [line, line width=0.15, <->, >=stealth] ([xshift=2em]DRP.south) -- (house5);
\path [line, line width=0.15, <->, >=stealth] ([xshift=2em]DRP.south) -- (house6);
\path [line, line width=0.15, <->, >=stealth] ([xshift=2em]DRP.south) -- (house7);
\path [line, line width=0.15, <->, >=stealth] ([xshift=2em]DRP.south) -- (house8);
\path [line, line width=0.15, <->, >=stealth] ([xshift=2em]DRP.south) -- (house9);
\path [line, line width=0.15, <->, >=stealth] ([xshift=2em]DRP.south) -- (house9);

\draw[rounded corners=2pt, fill=blue, opacity=0.2]
  (0.25,-3.2) rectangle ++(1.7,0.7) node[pos=0.85, opacity=1.0, color=blue] {\scriptsize{\textbf{PDR}}};
  
\path [line, line width=0.15, <->, >=stealth] (house9.east) -- (Ut);

\path [line, line width=0.15, <->, >=stealth] (Ut) -- (SC); 
\path [line, line width=0.15, <->, >=stealth] (WSM) -- (SC);
\path [line, line width=0.15, <->, >=stealth] (Ut.west) -- (DRP); 

\path [line, line width=0.15, <->, >=stealth] (Ut.north west) -- (WSM);
  
\end{tikzpicture}

\caption{Interactions of Agents in Residential DR}
\label{fig:DR_Chart}
\vspace*{-0.2cm}
\end{figure}

In 2015, the California Public Utilities Commission launched a Demand Response Auction Mechanism (DRAM) \cite{State-of-California-CPUC:2015aa}, requiring electric utilities to procure a certain amount of reduction capacity from DR providers. These aggregators incentivize their customers (also called ``Proxy Demand Resource'' (PDR) \cite{CAISO:2014aa}) under contract to temporarily reduce their consumption relative to their projected usage without intervention, referred to in this context as counterfactual or \textit{baseline}, based on which compensations for (non-)fulfilled reductions are settled: If the consumer uses less (more) energy than the baseline, she receives a reward (incurs a penalty). Figure \ref{fig:DR_Chart} illustrates the interactions between agents.

The estimation of the materialized reduction arguably is the most critical component of the DR bidding process. If the reductions are estimated with a biased counterfactual, either the DR provider or the utility clearing the bids is systematically discriminated against. If the baseline is unbiased but plagued by high variance, the profit settlement is highly volatile. Existing baselines employed by major power grid operators in the United States (e.g. California Independent System Operator (CAISO), New York ISO) are calculated with simple arithmetic averages of previous observations \cite{CAISO:2014aa} and therefore are inaccurate. The estimation of more accurate baselines is a significant contribution of this paper.

\subsection{Contributions}
We estimate the average treatment effect (ATE) of hour-ahead notifications on the reduction of electricity consumption by evaluating a Randomized Controlled Trial (RCT) on $\approx$ 7,000 residential households in California serviced by the three main electric utilities (PG\&E, SDG\&E, SCE), using funds provided by the California Energy Commission.

We estimate an ATE of $-0.10$ kWh per DR Event and user and discover noticeable geographic and temporal heterogeneity among users, as the largest estimated reductions occur in summer months as well as in regions with warmer climate.

Further, we utilize household census data to detect treatment effect heterogeneity of users by modifying the training procedure of classification and decision trees \cite{Breiman:1984aa} in a way that partitions the feature space into axis-aligned regions with piecewise constant treatment effects. The splits at each interior node are optimized by penalizing differences in covariates and rewarding differences of the mean outcome between treated and control units. This approach is inspired by \cite{Athey:2016aa}, but departs from the original model to fit our setting. Additionally, we benchmark its performance to a more naive $k$-means clustering approach that simply fits centroids in the feature space, each of which corresponds to a particular ``group'' with the identical treatment effect.

\subsection{Background and Related Work}
Performing causal inference to estimate the effect of an intervention on a metric of interest has long been tantamount to setting up a randomized controlled trial (RCT), whereby the difference in means between the control and treatment group is defined to be the average treatment effect. However, many experiments are infeasible or critical due to budget and ethical concerns. With the rapid growth of collected user data, a new direction of research at the intersection of machine learning and economics is concerned with using nonexperimental estimation techniques in lieu of RCTs. The general idea of such nonexperimental estimation techniques is to partition observations under treatment and control in order to fit a nominal model on the latter set, which, when applied on the treatment set, yields the treatment effect of interest.

Examples for such models are \cite{Bollinger:2015aa}, who evaluates welfare effects of home automation by calculating the Kolmogorov-Smirnov Statistic between users, or \cite{Abadie:2012aa}, who constructs a convex combination of US states as the counterfactual estimate for tobacco consumption to estimate the effect of a tobacco control program in California on tobacco consumption.

Our setting involves time-series prediction of smart meter data, which is essentially a short-term load forecasting (STLF) problem. Within STLF, tools employed are ARIMA models with a seasonal component \cite{Taylor:2007aa} and classic regression models where support vector regression (SVR) and neural networks yield the highest accuracy \cite{Senjyu:2002aa, Elattar:2010aa}. For a comprehensive comparison of ML techniques for forecasting and differing levels of load aggregation, see \cite{Mirowski:2014aa}.

In our previous work \cite{Zhou:2016aa, Zhou:2016ab, Zhou:2018ab}, we have developed de-biased nonexperimental estimators for predicting individual treatment effects (ITEs) of a residential DR program and benchmarked their accuracy against an experimental gold standard. We achieved this by modifying unit-level fixed effects, which are often used in fixed effects regressions in applied economics \cite{Allcott:2011aa}, into explicit user-level terms, making the resulting regression model amenable to general ML models.

While we begin by revising the concepts of nonexperimental estimators and an overview of the experimental setting, the focus of this paper lies in more advanced analyses using $K$-means clustering and causal decision trees to explore heterogeneity of treatment effects across the population based on demographic features. The idea of pursuing this direction is well-motivated from a practical perspective, as detecting heterogeneous subpopulations would allow us to design adaptive targeting schemes that customize incentives that seek to maximize a certain objective, e.g. cost efficiency. 

%
%

\section{Experimental Setup}
\label{sec:Experiment_Data}

\subsection{Setup of the Experiment}
\begin{figure}[h]
\vspace*{-0.32cm}
\centering
\begin{tikzpicture}
\node [status, align=left, inner sep=2pt, minimum size=1pt] (All) {\small{All Participants}};

\node [status, below=1.3em of All, xshift=-2cm, inner sep=2pt, minimum size=1pt] (Enc) {\small{\begin{tabular}{l}Treatment$-$\\Encouraged\end{tabular}}};
\node [status, below=1.3em of All, xshift=1.0cm, inner sep=2pt, minimum size=1pt] (Non-Enc) {\small{\begin{tabular}{l}Treatment$-$\\Non-Encouraged\end{tabular}}};
\node [status, below=1.7em of All, xshift=4.5cm, inner sep=2pt, minimum size=1pt] (Ctrl) {\small{Control}};
\node [status, draw=none, below=1.7em of Enc, xshift=1.5cm, inner sep=2pt, minimum size=1pt] (Estim) {\small{Estimate Responses}};
\node [status, below=1.7em of Estim, xshift=-1.5cm, inner sep=2pt, minimum size=1pt] (tgthigh) {\small{Targeted-High}};
\node [status, below=1.7em of Estim, xshift=1.0cm, inner sep=2pt, minimum size=1pt] (tgtlow) {\small{Targeted-Low}};
\node [status, below=1.7em of Estim, xshift=3.5cm, inner sep=2pt, minimum size=1pt] (nontgt) {\small{Non-Targeted}};
\node [status, draw=none, below=1.8em of tgthigh, xshift=1.5cm, inner sep=2pt, minimum size=1pt] (shuffle) {\small{Shuffle Users}};
\node [status, below=1.7em of shuffle, xshift=-1.5cm, inner sep=2pt, minimum size=1pt] (msuasion) {\small{Moral Suasion}};
\node [status, below=1.7em of shuffle, xshift=0.6cm, inner sep=2pt, minimum size=1pt] (price) {\small{Price}};
\node [status, below=1.7em of shuffle, xshift=2.9cm, inner sep=2pt, minimum size=1pt] (suasionprice) {\small{Moral Suasion \& Price}};
\node [status, below=1.7em of shuffle, xshift=5.1cm, inner sep=2pt, minimum size=1pt] (ph3control) {\small{Control}};

\draw[->, >=stealth] (All.south) -- node [near start] {} +(0,-0.2) -| (Enc) node [above, yshift=-2cm, xshift=0.2cm] {};
\draw[->, >=stealth] (All.south) -- node [near start] {} +(0,-0.2) -| (Non-Enc) node [above, yshift=-2cm, xshift=0.2cm] {};
\draw[->, >=stealth] (All.south) -- node [near start] {} +(0,-0.2) -| (Ctrl) node [above, yshift=-2cm, xshift=0.2cm] {};

\draw[->, >=stealth] (Enc.south) -- node [near start] {} +(0,-0.2) -| (Estim) node [above, yshift=-2cm, xshift=0.2cm] {};
\draw[->, >=stealth] (Non-Enc.south) -- node [near start] {} +(0,-0.2) -| (Estim) node [above, yshift=-2cm, xshift=0.2cm] {};

\draw[->, >=stealth] (Estim.south) -- node [near start] {} +(0,-0.2) -| (tgthigh) node [above, yshift=-2cm, xshift=0.2cm] {};
\draw[->, >=stealth] (Estim.south) -- node [near start] {} +(0,-0.2) -| (tgtlow) node [above, yshift=-2cm, xshift=0.2cm] {};
\draw[->, >=stealth] (Estim.south) -- node [near start] {} +(0,-0.2) -| (nontgt) node [above, yshift=-2cm, xshift=0.2cm] {};

\draw[->, >=stealth] (tgthigh) -- node [near start] {} +(0,-0.35) -| (shuffle) node [above, yshift=-2cm, xshift=0.2cm] {};
\draw[->, >=stealth] (tgtlow) -- node [near start] {} +(0,-0.35) -| (shuffle) node [above, yshift=-2cm, xshift=0.2cm] {};
\draw[->, >=stealth] (nontgt) -- node [near start] {} +(0,-0.35) -| (shuffle) node [above, yshift=-2cm, xshift=0.2cm] {};
\draw[->, >=stealth] (Ctrl) -- node [near start] {} +(0,-2.73) -| (shuffle) node [above, yshift=-2cm, xshift=0.2cm] {};

\draw[->, >=stealth] (shuffle.south) -- node [near start] {} +(0,-0.2) -| (msuasion) node [above, yshift=-2cm, xshift=0.2cm] {};
\draw[->, >=stealth] (shuffle.south) -- node [near start] {} +(0,-0.2) -| (price) node [above, yshift=-2cm, xshift=0.2cm] {};
\draw[->, >=stealth] (shuffle.south) -- node [near start] {} +(0,-0.2) -| (suasionprice) node [above, yshift=-2cm, xshift=0.2cm] {};
\draw[->, >=stealth] (shuffle.south) -- node [near start] {} +(0,-0.2) -| (ph3control) node [above, yshift=-2cm, xshift=0.2cm] {};

\draw[black, dashed, color=blue, rounded corners=2pt, line width=1.5] ($(Enc.north west)+(-0.2,0.35)$)  rectangle ($(Non-Enc.south east)+(2.85,-0.4)$) node [above, xshift=-5.45em, yshift=-0.18em] {\small{Phase 1 (90 days)}};

\draw[black, dashed, color=mygreen, rounded corners=2pt, line width=1.5] ($(tgthigh.north west)+(-0.25,0.25)$)  rectangle ($(nontgt.south east)+(0.25,-0.4)$) node [above, xshift=-3.45em, yshift=-1.61em] {\small{Phase 2 (90 days)}};

\draw[black, dashed, color=red, rounded corners=2pt, line width=1.5] ($(msuasion.north west)+(-0.05,0.25)$)  rectangle ($(ph3control.south east)+(0.05,-0.2)$) node [above, xshift=-3.15em, yshift=-1.61em] {\small{Phase 3 (90 days)}};

\end{tikzpicture}
\vspace*{-1.35cm}
\caption{Setup of Experiment}
\vspace*{-0.02cm}
\label{fig:setup_experiment}
\end{figure}
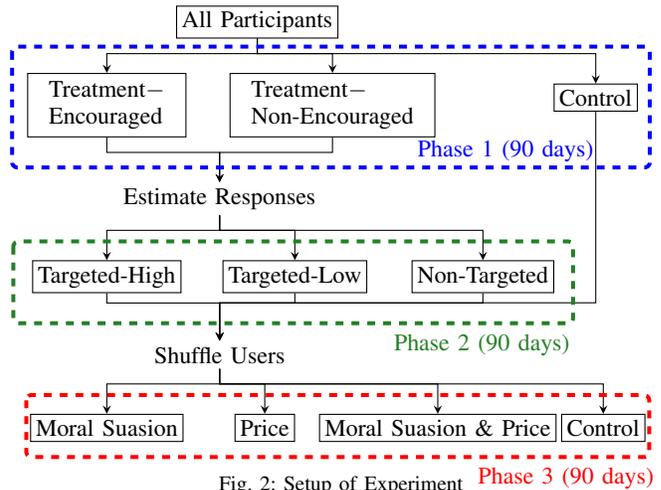

The experiment is carried out by OhmConnect, Inc., using funds provided by the California Energy Commission. Figure \ref{fig:setup_experiment} draws a flowchart of the experimental setup.

Over the course of the experimental time period (Nov. 2016 - Dec. 2017), each consumer that signs up for the study is \textit{randomly} assigned to one of the following groups:
\begin{itemize}
\item Treatment$-$Encouraged: The user receives an average number of 25 DR events in the 90 days following recruitment, with event incentive levels selected uniformly at random from the set $\mathcal{R} = \lbrace 0.05, 0.25, 0.50, 1.00, 3.00\rbrace \frac{\text{USD}}{\text{kWh}}$. Also, the user is given a rebate for purchasing a smart home automation device.
\item Treatment$-$Non-Encouraged: Same as in Treatment-Encouraged, but without smart home automation rebate.
\item Control (Initially Delayed): Users do not receive any notifications for DR events for a period of 90 days after enrollment.
\end{itemize}
These three groups form Phase 1 of the experiment. 
Users in Phase 1 treatment groups that have reached 90 days of age are pooled and assigned to one of three different groups for Phase 2 interventions. Users receive either large incentives, smaller ones, or a mix of both depending on the group they are assigned to. Lastly, users with completed Phase 2 as well as Phase 1 control users after 90 days of age undergo Phase 3 (moral suasion), which occurs on an event-by-event level and selectively includes moral suasion in text messages.

In Sections \ref{sec:nonexperimental_treatment_effect_estimation} and \ref{sec:Nonexperimental_Estimation_Results}, we evaluate Phase 1 of the experiment. Phase 2 and 3 are analyzed in our previous work and are hence omitted in this paper. Instead, we focus on detecting heterogeneity in treatment effects in Sections \ref{sec:causal_trees} and \ref{sec:kmeans_clustering}.

\section{Nonexperimental Treatment Effect Estimation}\label{sec:nonexperimental_treatment_effect_estimation}
\subsection{Potential Outcomes Framework}\label{sec:potential_outcomes_framework}
To estimate the effect of the DR intervention program, we adopt the \textit{potential outcomes} framework introduced by Rubin (1974) \cite{Rubin:1974aa}. Let $\mathcal{I} = \lbrace 1, \ldots, n\rbrace$ denote the set of users. The indicator $D_{it} \in \lbrace 0, 1\rbrace$ encodes the fact whether or not user $i$ received DR treatment at time $t$. Each user is equipped with a consumption time series $\mathbf{y}_i = \lbrace y_{i1}, \ldots, y_{i\tau} \rbrace$ and associated covariates $X_i = \lbrace \mathbf{x}_{i1}, \ldots, \mathbf{x}_{i\tau} \rbrace~\in\times_{i=1}^\tau\mathcal{X}_i$, $\mathcal{X}_i\subset\mathbb{R}^{n_x}$, where time is indexed by $t\in \mathbb{T} = \lbrace 1, \ldots, \tau \rbrace$ and $n_x$ is the dimension of the covariate space $\mathcal{X}_i$. Let $y_{it}^0$ and $y_{it}^1$ denote user $i$'s electricity consumption at time $t$ for $D_{it}=0$ and $D_{it}=1$, respectively. Let $\mathcal{C}_i$ and $\mathcal{T}_i$ denote the set of control and treatment times for user $i$. That is,
\begin{align}
\mathcal{C}_i = \lbrace t\in \mathbb{T}~|~D_{it} = 0 \rbrace,\quad\mathcal{T}_i = \lbrace t\in \mathbb{T}~|~D_{it} = 1 \rbrace.
\end{align}
The number of treatment hours is much smaller than the number of non-treatment hours. Thus $0 < |\mathcal{T}_i|/|\mathcal{C}_i| \ll 1$.

Further, let $\mathcal{D}_{i,t}$ and $\mathcal{D}_{i,c}$ denote user $i$'s covariate-outcome pairs of treatment and control times, respectively. That is,
\begin{align}
\hspace*{-0.26cm}\mathcal{D}_{i,t} = \lbrace (\mathbf{x}_{it}, y_{it})~|~t\in\mathcal{T}_i \rbrace,~\mathcal{D}_{i,c} = \lbrace (\mathbf{x}_{it}, y_{it})~|~t\in\mathcal{C}_i \rbrace.
\end{align}
The one-sample estimate of the treatment effect on user $i$ at time $t$, given the covariates $\mathbf{x}_{it}\in\mathbb{R}^{n_x}$, is
\begin{align}\label{eq:one_sample_estimate}
\beta_{it}(\mathbf{x}_{it}) := y_{it}^1(\mathbf{x}_{it}) - y_{it}^0(\mathbf{x}_{it})\quad \forall~i\in\mathcal{I},~t\in\mathbb{T},
\end{align}
which varies across time, the covariate space, and the user population. Marginalizing this one-sample estimate over the set of treatment times $\mathcal{T}_i$ and the covariate space $\mathcal{X}_i$ yields the user-specific Individual Treatment Effect (ITE) $\beta_i$
\begin{align}\label{eq:individual_treatment_effect}
\beta_i := \mathbb{E}_{\mathcal{X}_i}\mathbb{E}_{t\in\mathcal{T}_i}\left[y_{it}^1 - y_{it}^0~\Big|~\mathbf{x}_{it}\right] = \frac{1}{|\mathcal{T}_i|}\sum_{t\in\mathcal{T}_i} (y_{it}^1 - y_{it}^0).
\end{align}
The average treatment effect on the treated (ATT) follows from \eqref{eq:individual_treatment_effect}:
\begin{align}\label{eq:ATT}
\text{ATT} := \mathbb{E}_{i\in\mathcal{I}}[\beta_i] = \frac{1}{|\mathcal{I}|}\sum_{i\in\mathcal{I}}\frac{1}{|\mathcal{T}_i|}\sum_{t\in\mathcal{T}_i}(y_{it}^1 - y_{it}^0).
\end{align}
Since users were put into different experimental groups in a \textit{randomized} fashion, the ATT and the average treatment effect (ATE) are identical \cite{Pischke:2009aa}.
Lastly, the conditional average treatment effect (CATE) on $\tilde{\mathbf{x}}$ is obtained by marginalizing the conditional distribution of one-sample estimates \eqref{eq:one_sample_estimate} on $\tilde{\mathbf{x}}$ over all users and treatment times, where $\tilde{\mathbf{x}}\in\mathbb{R}^{\tilde{n}_x}$ is a subvector of $\mathbf{x}\in\mathbb{R}^{n_x},~0 < \tilde{n}_x < n_x$:
\begin{align}\label{eq:CATT}
\text{CATE}(\tilde{\mathbf{x}}) := \mathbb{E}_{i\in\mathcal{I}}\mathbb{E}_{t\in\mathcal{T}_i}\left[(y_{it}^1 - y_{it}^0)~\Big|~\tilde{\mathbf{x}}_{it}=\tilde{\mathbf{x}}\right].
\end{align}
The CATE captures heterogeneity among users, e.g. with respect to specific hours of the day, the geographic distribution of users, the extent to which a user possesses ``smart home'' appliances, group or peer effects, etc.
To rule out the existence of unobserved factors that could influence the assignment mechanism generating the complete observed data set $\lbrace (\mathbf{x}_{it}, y_{it}, D_{it})~|~i\in\mathcal{I}, t\in\mathbb{T} \rbrace$, we make the following standard assumptions:
\begin{assumption}[Unconfoundedness of Treatment Assignment]\label{as:ignorable_treatment_assignment} Given the covariates $\lbrace \mathbf{x}_{it}\rbrace
_{t\in\mathbb{T}}$, the potential outcomes are independent of treatment assignment:
\begin{align}
(y_{it}^0, y_{it}^1) \perp D_{it}~|~\mathbf{x}_{it}\quad\forall i\in\mathcal{I}, t\in\mathbb{T}.
\end{align}
\end{assumption}
\begin{assumption}[Stationarity of Potential Outcomes]\label{as:stationarity_potential_outcomes} Given the covariates $\lbrace \mathbf{x}_{it}\rbrace
_{t\in\mathbb{T}}$, the potential outcomes are independent of time, that is,
\begin{align}
(y_{it}^0, y_{it}^1) \perp t~|~\mathbf{x}_{it}\quad\forall i\in\mathcal{I}, t\in\mathbb{T}.
\end{align}
\end{assumption}
Assumption \ref{as:ignorable_treatment_assignment} is the ``ignorable treatment assignment'' assumption introduced by Rosenbaum and Rubin \cite{Rosenbaum:1983aa}. Under this assumption, the assignment of DR treatment to users is implemented in a \textit{randomized} fashion, which allows the calculation of unbiased ATEs \eqref{eq:ATT} and CATEs \eqref{eq:CATT}. Assumption \ref{as:stationarity_potential_outcomes}, motivated by the time-series nature of the observational data, ensures that the set of observable covariates $\lbrace\mathbf{x}_{it}~|~t\in\mathbb{T}\rbrace$ can capture seasonality effects in the estimation of the potential outcomes. That is, the conditional distribution of the potential outcomes, given covariates, remains constant.


The \textit{fundamental problem of causal inference} \cite{Holland:1986aa} refers to the fact that either the treatment or the control outcome can be observed, but never both (granted there are no missing observations). That is,
\begin{align}
y_{it} = y_{it}^0 + D_{it}\cdot(y_{it}^1 - y_{it}^0)\quad\forall~t\in\mathbb{T}.
\end{align}

Thus, the ITE \eqref{eq:individual_treatment_effect} is not identified, because one and only one of both potential outcomes is observed, namely $\lbrace y_{it}^1~|~t\in\mathcal{T}_i \rbrace$ for the treatment times and $\lbrace y_{it}^0~|~t\in\mathcal{C}_i\rbrace$ for the control times. It therefore becomes necessary to estimate counterfactuals. 

\subsection{Non-Experimental Estimation of Counterfactuals}\label{eq:nonexperimental_estimation_counterfactuals}
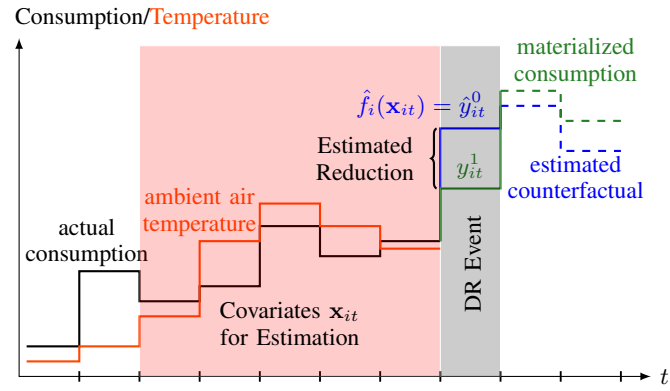
\begin{figure}[h]
\vspace*{-0.3cm}
\centering
\hspace*{-0.3cm}
\begin{tikzpicture}[>=latex, y=2cm, font=\small]
    \draw [<->] (0,2.25) node [above, xshift=1.64cm] {Consumption/{\color{myorange}{Temperature}}} |- (8.4,0) node [right] {$t$};
    \foreach \x in {1,...,10}{
   \draw [line width=0.25mm](\x/1.25,2pt) -- (\x/1.25,-2pt);
}

\draw (0.8,0) node[below=3pt] {\scriptsize{8:00}};
\draw (1.6,0) node[below=3pt] {\scriptsize{9:00}};
\draw (2.4,0) node[below=3pt] {\scriptsize{10:00}};
\draw (3.2,0) node[below=3pt] {\scriptsize{11:00}};
\draw (4.0,0) node[below=3pt] {\scriptsize{12:00}};
\draw (4.8,0) node[below=3pt] {\scriptsize{13:00}};
\draw (5.6,0) node[below=3pt] {\scriptsize{14:00}};
\draw (6.4,0) node[below=3pt] {\scriptsize{15:00}};
\draw (7.2,0) node[below=3pt] {\scriptsize{16:00}};
\draw (8.0,0) node[below=3pt] {\scriptsize{17:00}};

\draw [black, thick] (0.1,0.2) -| (0.8,0.7) node [above, xshift=0.1cm, yshift=-0.1cm] {\small{\begin{tabular}{c}actual\\consumption\end{tabular}}} -| (1.6,0.5) -| (2.4,0.6) -| (3.2,1.0) -| (4.0,0.8) -| (4.8,0.9) -- (5.6, 0.9);

\draw [myorange, thick] (0.1, 0.1) -| (0.8, 0.2) -| (1.6, 0.4) node [above, xshift=0.8cm, yshift=0.9cm] {\begin{tabular}{c}ambient air\\temperature\end{tabular}} -| (2.4, 0.9) -| (3.2, 1.15) -| (4, 1.0) -| (4.8, 0.85) -- (5.6, 0.85);

\draw[white, fill=red, opacity=0.2, line width=0.1] (1.6,0) rectangle (5.6, 2.2);
\node[status, draw=none] at (3.6,0.6) {\begin{tabular}{c}Covariates $\mathbf{x}_{it}$\\for Estimation\end{tabular}};

\draw [blue, thick] (5.598, 0.9) -- (5.598, 1.65) node [above, xshift=-0.25cm] {$\hat{f}_i(\mathbf{x}_{it}) = \hat{y}_{it}^0$} -| (6.4, 1.8) -- (6.4, 1.8);
\draw [blue, dashed, thick] (6.4, 1.8) -| (7.2, 1.5) node [below, xshift=0.2cm, yshift=0.1cm] {\begin{tabular}{c}estimated\\counterfactual\end{tabular}} -| (8, 1.5);

\draw [mygreen, thick] (5.6, 0.9) -- (5.6,1.25) node [above, xshift=0.4cm] {$y_{it}^1$} -| (6.4,1.9) -- (6.4,1.9);
\draw [mygreen, dashed, thick] (6.4,1.9) -| (7.2,1.7) node [above, xshift=0.2cm, yshift=0.3cm] {\begin{tabular}{c}materialized\\consumption\end{tabular}} -| (8,1.7);

\draw[white, fill=black, opacity=0.2, line width=0.1] (5.6,0) rectangle (6.4, 2.2);
\node[status, draw=none, rotate=90] at (5.5,0.75) {DR Event};

\node[status, draw=none] at (4.6,1.7) {\begin{tabular}{c}Estimated\\Reduction\end{tabular}};

\draw [thick, decorate,decoration={brace,amplitude=2pt}] (5.55,1.25) -- (5.55,1.65);
\end{tikzpicture}
\caption{Estimation of the Counterfactual $\hat{y}_{it}^0$ using Treatment Covariates $\mathbf{x}_{it}$ and Predicted Reduction $y_{it}^1 - \hat{y}_{it}^0$}
\label{fig:counterfactual_estimation}
\vspace*{-0.2cm}
\end{figure}

Consider the following model for the estimation of counterfactuals:
\begin{align}\label{eq:observation_outcome_model}
y_{it} = f_i(\mathbf{x}_{it}) + D_{it}\cdot\beta_{it}(\mathbf{x}_{it}) + \varepsilon_{it},
\end{align}
where $\varepsilon_{it}$ denotes noise uncorrelated with covariates and treatment assignment. $f_i(\cdot):\mathbb{R}^{n_x}\mapsto \mathbb{R}$ is the conditional mean function and pertains to $D_{it} = 0$. To obtain an estimate for $f_i(\cdot)$, denoted with $\hat{f}_i(\cdot)$, control outcomes $\lbrace y_{it}^0~|~t\in\mathcal{C}_i\rbrace$ are first regressed on $\lbrace \mathbf{x}_{it}~|~t\in\mathcal{C}_i\rbrace$, namely their observable covariates. In a second step, the counterfactual $\hat{y}_{it}^0$ for any $t\in\mathcal{T}_i$ can be estimated by evaluating $\hat{f}_i(\cdot)$ on its associated covariate vector $\mathbf{x}_{it}$. Finally, subtracting $\hat{y}_{it}^0$ from $y_{it}^1$ isolates the one-sample estimate $\beta_{it}(\mathbf{x}_{it})$, from which the user-specific ITE \eqref{eq:individual_treatment_effect} can be estimated. Figure \ref{fig:counterfactual_estimation} illustrates this process of estimating the reduction during a DR event by subtracting the actual consumption $y^1_{it}$ from the predicted counterfactual $\hat{y}_{it}^0 = \hat{f}_i(\mathbf{x}_{it})$. 
We restrict our estimators $f_i(\cdot)$ to a single hour prediction horizon as DR events are at most one hour long.

To estimate $f_i(\cdot)$, we use the following classical regression methods \cite{Hastie:2009aa}, referred to as \textit{estimators}:
\begin{itemize}\compresslist
\item (E1): Ordinary Least Squares Regression (OLS) 
\item (E2): L1 Regularized (LASSO) Linear Regression (L1)
\item (E3): L2 Regularized (Ridge) Linear Regression (L2)
\item (E4): $k$-Nearest Neighbors Regression (KNN)
\item (E5): Decision Tree Regression (DT)
\item (E6): Random Forest Regression (RF)
\end{itemize}
DT (E5) and RF (E6) follow the procedure of Classification and Regression Trees \cite{Breiman:1984aa}. We compare estimators (E1)-(E6) to the CAISO 10-in-10 Baseline (BL) \cite{CAISO:2014aa}, 
which, for any given hour on a weekday, is calculated as the mean of the hourly consumptions on the 10 most recent business days during the selected hour. For weekend days and holidays, the mean of the 4 most recent observations is calculated. This BL is further adjusted with a \textit{Load Point Adjustment}, which corrects the BL by a factor proportional to the consumption three hours prior to a DR event \cite{CAISO:2014aa}.
To estimate user $i$'s counterfactual outcome $\hat{y}_{it}^0$ during a DR event $t\in\mathcal{T}_i$, we use the following covariates:
\begin{itemize}
\item 5 hourly consumption values preceding time $t$
\item Air temperature at time $t$ and 4 preceding measurements
\item Hour of the day, an indicator variable for (non-)business days, and month of the year as categorical variables
\end{itemize}
Thus, the covariate vector writes
\begin{equation}\label{eq:covariate_vector_nonexperimental}
\begin{aligned}
\mathbf{x}_{it} = [&y_{it-1}^0\quad \cdots\quad y_{it-5}^0 \quad T_{it}\quad \cdots\quad T_{it-4}\\
&\mathrm{C}(\mathrm{HoD}_{it}):\mathrm{C}(\mathrm{is\textunderscore Bday}_{it}) \quad \mathrm{C}(\mathrm{MoY}_{it})].
\end{aligned}
\end{equation}
In \eqref{eq:covariate_vector_nonexperimental}, $T_{it}$ denotes temperature, $\mathrm{HoD}_{it}$ hour of day, $\mathrm{is\textunderscore Bday}_{it}$ an indicator for business days, and $\mathrm{MoY}_{it}$ the month of year for user $i$ at time $t$. ``C'' denotes categorical variables and ``:'' their interaction.

\subsection{Estimation of Individual Treatment Effects}\label{sec:estimation_ITEs}
To obtain point estimates for user $i$'s ITE $\beta_i$, we simply average all one-sample estimates \eqref{eq:one_sample_estimate} according to \eqref{eq:individual_treatment_effect}. 
To obtain an estimate of whether or not a given user $i$ has actually reduced consumption, we utilize a nonparametric permutation test with the null hypothesis of a zero ITE:
\begin{align}
H_0: \beta_i=0, \quad \quad H_1: \beta_i\neq 0.\label{eq:null_hyp_permutation}
\end{align}

Given user $i$'s paired samples $\lbrace z_{it} = \hat{y}_{it}^0 - y_{it}^1~|~t\in\mathcal{T}_i\rbrace$ during DR periods, the $p$-value associated with $H_0$ \eqref{eq:null_hyp_permutation} is 
\begin{align}\label{eq:p_value_permutation_test}
p = \frac{\sum_{D\in\mathcal{P}_i}\mathds{1}(\bar{D} \leq \hat{\beta}_i)}{2^{|\mathcal{T}_i|}}.
\end{align}
In \eqref{eq:p_value_permutation_test}, $\bar{D}$ denotes the mean of $D$. $\mathcal{P}_i$ denotes the set of all possible assignments of signs to the pairwise differences in the set $\lbrace z_{it} = y_{it}^1 - \hat{y}_{it}^0 ~|~t\in\mathcal{T}_i\rbrace$. That is,
\begin{align}
\hspace*{-0.17cm}\mathcal{P}_i = \lbrace s_1 z_{i1}, \ldots, s_{|\mathcal{T}_i|}z_{i|\mathcal{T}_i|}~|~ s_j \in\lbrace \pm 1\rbrace, 1\leq j\leq |\mathcal{T}_i|\rbrace
\end{align}
which is of size $2^{|\mathcal{T}_i|}$. Finally, the $p$-value from \eqref{eq:null_hyp_permutation} is calculated as the fraction of all possible assignments whose means are less than or equal the estimated ITE $\hat{\beta}_i$. In practice, as the number of DR events per user in Phase 1 is about 25 (see Figure \ref{fig:setup_experiment}), the number of total possible assignments becomes computationally infeasible. Thus, we randomly generate a subset of $10^5$ assignments from $\mathcal{P}_i$ to compute the $p$-value in \eqref{eq:p_value_permutation_test}. Moreover, we use the percentile bootstrap method \cite{Efron:1994aa} to compute a confidence interval of the estimated ITE for user $i$ around the point estimate $\hat{\beta}_i$.

%
%


\section{Nonexperimental Estimation Results}\label{sec:Nonexperimental_Estimation_Results}

\subsection{Average Treatment Effect}
\begin{figure}[hbtp]
\vspace*{-0.35cm}
\centering
\includegraphics[width=0.49\textwidth]{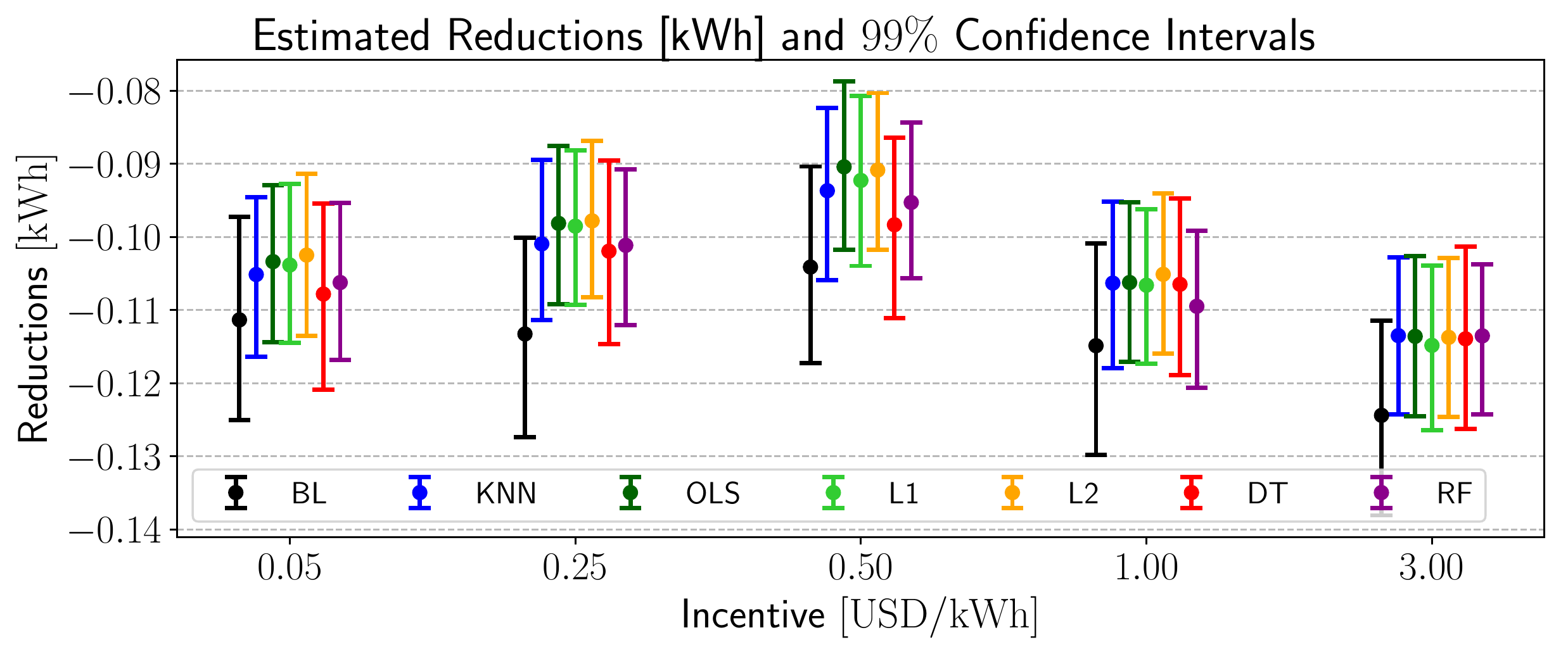}
\caption{CATEs by Incentive Level \& Confidence Intervals}
\label{fig:CATE_Incentive_Level}
\vspace*{-0.12cm}
\end{figure}

Figure \ref{fig:CATE_Incentive_Level} shows ATE point estimates and their 99\% bootstrapped confidence intervals conditional on differing reward levels for all estimators as well as the CAISO BL. Due to the empirical de-biasing procedure (c.f. \cite{Zhou:2016aa, Zhou:2016ab, Zhou:2018ab}), the point estimates for estimators (E1)-(E6) are close to each other. BL appears to be biased in favor of the users, as it systematically predicts larger reductions than (E1)-(E6).

The ATE averaged over the predictions of estimators (E1)-(E6) is $-0.105$ kWh / $-11.5\%$. The intercept and the slope of the demand curve are $-0.099$ kWh / $-0.013$ kWh/USD, meaning that users reduce an additional 0.013 kWh per dollar offered, a small change. Due to the idiosyncratic nature of the CATE for $r=0.5\frac{\text{USD}}{\text{kWh}}$, the slope and intercept have to be interpreted with caution. However, the results give rise to a notable correlation between incentive levels and reductions.


To compare the prediction accuracy of the estimators, Table \ref{tab:CI_width_Incentive_Level} reports the width of the confidence intervals for each method and incentive level. The inferiority of the CAISO baseline compared to the non-experimental estimators, among which RF achieves the tightest confidence intervals, becomes apparent. Therefore, in the remainder of this paper, we restrict all results achieved with non-experimental estimators to those obtained with RF.

\begin{table}[hbtp]
\scriptsize
\vspace*{0.15cm}
\centering
\begin{tabular}{c|ccccc}
\toprule
\multicolumn{6}{c}{Width of CATE Confidence Intervals (kWh) by Incentive Level} \\
\hline
\\[-0.8em]
 & 0.05 $\frac{\text{USD}}{\text{kWh}}$ & 0.25 $\frac{\text{USD}}{\text{kWh}}$ & 0.50 $\frac{\text{USD}}{\text{kWh}}$ & 1.00 $\frac{\text{USD}}{\text{kWh}}$ & 3.00 $\frac{\text{USD}}{\text{kWh}}$ \\
[0.3em]
\hline
BL & $0.0277$ & $0.0273$ & $0.0269$ & $0.0289$ & $0.0266$\\
KNN & $0.0218$ & $0.0219$ & $0.0235$ & $0.0227$ & $0.0214$\\
OLS & $0.0214$ & $0.0217$ & $0.0230$ & $0.0219$ & $0.0219$\\
L1 & $0.0217$ & $0.0211$ & $0.0233$ & $0.0212$ & $0.0225$\\
L2 & $0.0221$ & $0.0214$ & $0.0214$ & $0.0219$ & $0.0218$\\
DT & $0.0255$ & $0.0251$ & $0.0247$ & $0.0241$ & $0.0250$\\
RF & $\textbf{0.0211}$ & $\textbf{0.0210}$ & $\textbf{0.0212}$ & $\textbf{0.0211}$ & $\textbf{0.0205}$\\
\bottomrule
\end{tabular}
\vspace{0.24cm}
\caption{Width of 99\% Confidence Intervals around ATE Point Estimate by Incentive Level, all Estimators}
\label{tab:CI_width_Incentive_Level}
\end{table}

\begin{figure}[hbtp]
\centering
\includegraphics[width=0.49\textwidth]{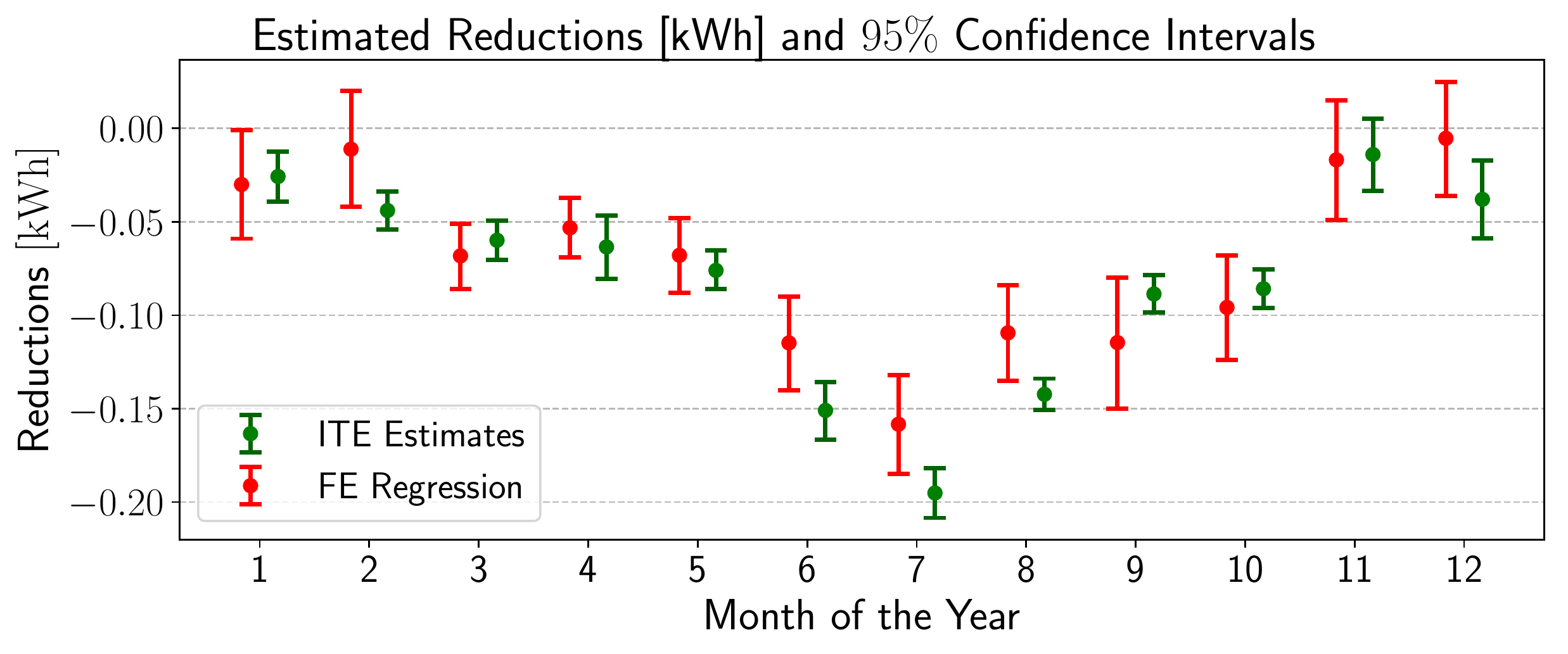}
\caption{CATE by Month, Fixed Effects Estimator (cf. \cite{Zhou:2018ab}) vs. Random Forest Estimator}
\label{fig:ATE_by_month}
\vspace*{-0.12cm}
\end{figure}

Figure \ref{fig:ATE_by_month} plots the CATE broken out by month of the year, where we compare the estimates from the Random Forest estimator to a Fixed Effects (FE) estimator, which serves as an experimental gold standard. For further explanation on this comparison, the reader is referred to our previous work \cite{Zhou:2018ab}. Here we focus our attention on the larger reduction in summer months which are presumably attributed to an increased usage of air conditioning.

\subsection{Individual Treatment Effects}
\begin{figure}[hbtp]
\vspace*{-0.0cm}
\centering
\includegraphics[width=0.49\textwidth]{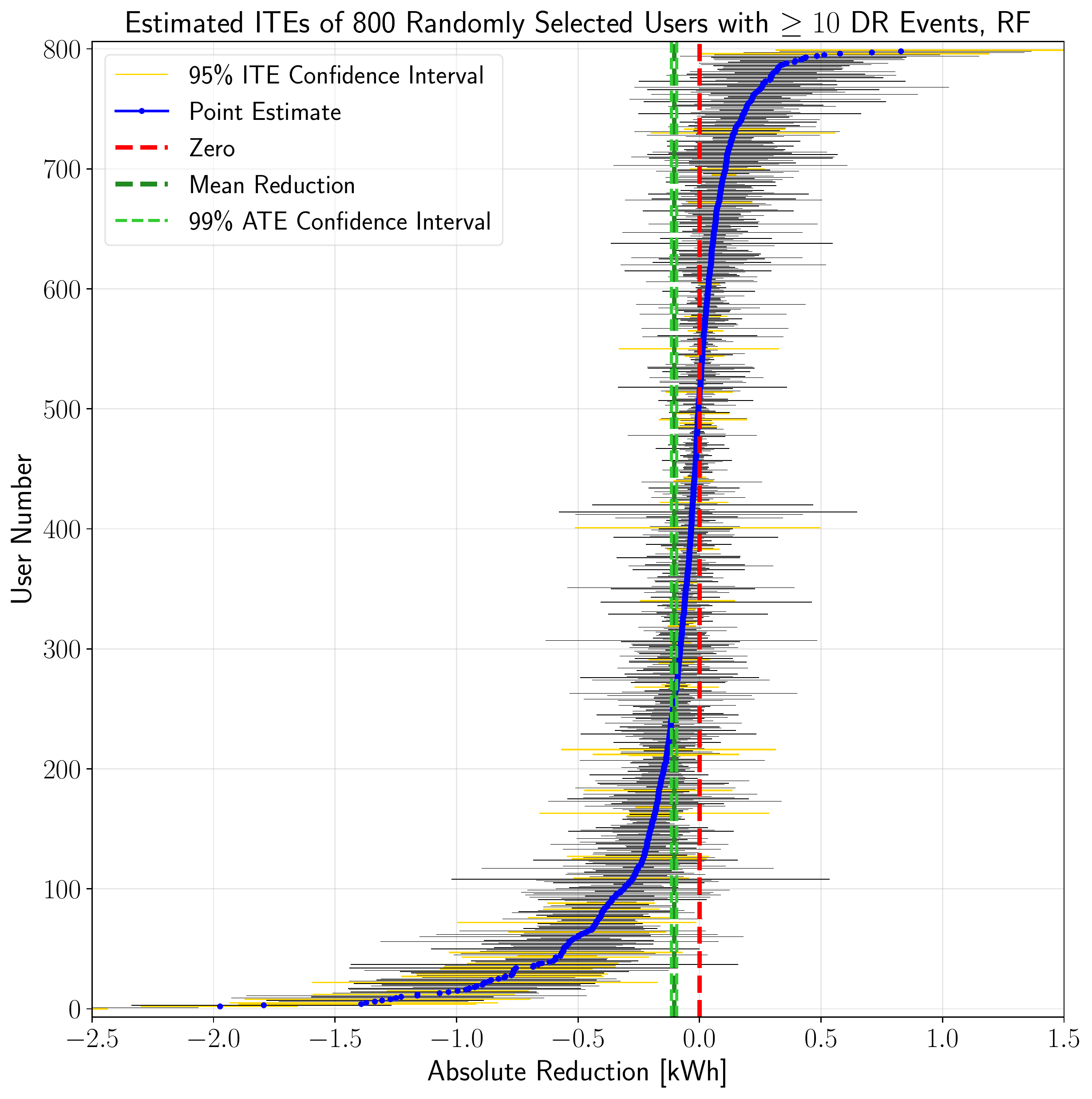}
\caption{Distribution of ITEs with Confidence Intervals}
\label{fig:ITEs_CIs}
\vspace*{-0.15cm}
\end{figure}

Figure \ref{fig:ITEs_CIs} plots ITEs for a randomly selected subset of 800 users who received at least 10 DR events in Phase 1, estimated with RF. Users are sorted by their point estimates (blue), whose 95\% bootstrapped confidence intervals are drawn in black. Yellow lines represent users with at least one active smart home automation device. By marginalizing the point estimates over all users with at least 10 events, we obtain an ATE of $-0.104$ kWh ($-$11.4\%), which is close to $-0.105$ kWh as reported earlier. The difference ensues from only considering users with at least 10 DR events. The 99\% ATE confidence interval is $[-0.115, -0.093]$ kWh.

Table \ref{tab:CATE_Automation_Status} reports estimated ATEs for users with or without active smart home automation devices, which are obtained by aggregating the relevant estimated ITEs from Figure \ref{fig:ITEs_CIs}. We notice larger responses as well as a larger percentage of estimated reducers among automated users.

\begin{table}[hbtp]
\scriptsize
\vspace*{-0.02cm}
\centering
\begin{tabular}{c|cccc}
\toprule
\multicolumn{5}{c}{ATEs Conditional on Automation Status for Users with $\geq$ 10 DR Events} \\
\hline
 & \# Users & \% Reducers & ATE (kWh) & ATE (\%) \\
\hline
Automated & $451$ & $79.2$ & $-0.279$ & $-36.7$ \\
Non-Automated & $4491$ & $63.6$ & $-0.087$ & $-9.62$ \\
All & $4942$ & $65.0$ & $-0.105$ & $-11.5$ \\
\bottomrule
\end{tabular}
\vspace{0.1cm}
\caption{Estimated CATEs by Automation Status, RF (E6)}
\label{tab:CATE_Automation_Status}
\vspace*{-0.1cm}
\end{table}

Table \ref{tab:percentage_reducers} reports the percentage of significant reducers for different confidence levels, obtained with the permutation test under the null \eqref{eq:null_hyp_permutation}. From Tables \ref{tab:CATE_Automation_Status} and \ref{tab:percentage_reducers}, it becomes clear that automated users show larger reductions than non-automated ones, which agrees with expectations.
\begin{table}[hbtp]
\scriptsize
\vspace*{0.15cm}
\centering
\begin{tabular}{c|ccc}
\toprule
\multicolumn{4}{c}{Fraction of Significant Reducers (among sample of size $4942$)} \\
\hline
 & $1-\alpha=0.9$ & $1-\alpha=0.95$ & $1-\alpha=0.99$ \\
\hline
\# Automated & $225$ & $205$ & $159$ \\
\% of Total & $49.9$ & $45.5$ & $35.3$ \\
\bottomrule
\# Non-Automated & $1382$ & $1162$ & $829$ \\
\% of Total & $30.8$ & $25.9$ & $18.5$ \\
\bottomrule
\# All & $1607$ & $1367$ & $988$ \\
\% of Total & $32.5$ & $27.7$ & $20.0$ \\
\bottomrule
\end{tabular}
\vspace{0.25cm}
\caption{Estimated Percentage of Significant Reducers according to Permutation Test, RF Estimator (E6)}
\label{tab:percentage_reducers}
\vspace*{-0.15cm}
\end{table}

\begin{figure}[hbtp]
\vspace*{-0.4cm}
\centering
\includegraphics[width=0.49\textwidth]{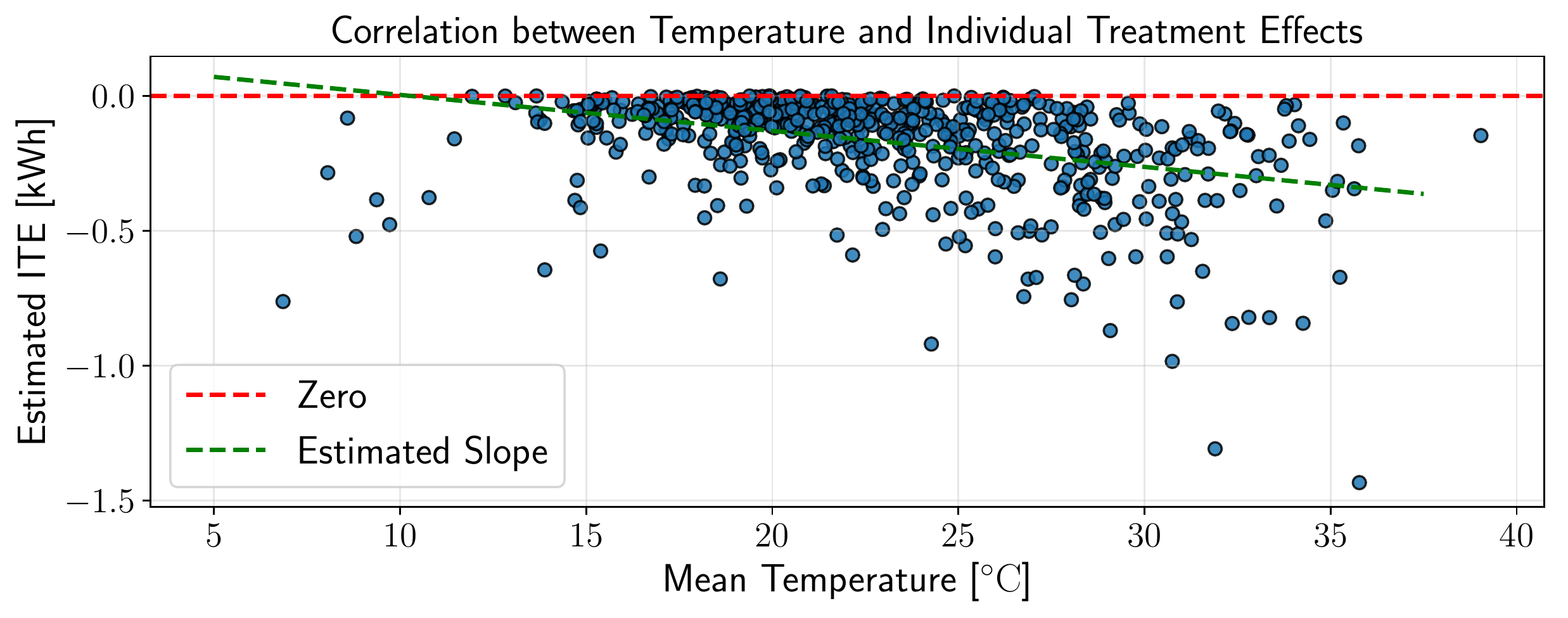}
\caption{Correlation between Average Ambient Air Temperature and ITEs.}
\label{fig:temperature_correlation_ITE}
\vspace*{-0.4cm}
\end{figure}

Larger reductions are estimated in warm summer months. To test the hypothesis whether or not there exists such a correlation, Figure \ref{fig:temperature_correlation_ITE} scatter plots estimated ITEs as a function of the average ambient air temperature observed during the relevant DR events. This gives rise to a noticeable positive correlation of ambient air temperature and the magnitude of reductions. Indeed, a subsequent hypothesis test with the null being a zero slope is rejected with a $p$-value of less than $1e-9$.


%
%

\section{Causal Decision Trees}\label{sec:causal_trees}
We utilize household census data to detect subpopulations with ``similar'' treatment effects. The data set consists of demographic features (age, ethnicity, relationship status, household size, income) as well as features that describe the electricity usage of customers, such as solar/electric/gas heating and the age of the house. Following best practices, we standardize the census data set (i.e. zero mean, unit variance).

\subsection{Training Algorithm}
We now explain the training procedure for fitting the causal tree. As a first step, we take all available training data, namely the collection of covariates, outcomes, and treatment indicators for all treated and control users over the entire experimental period, which we denote as $\lbrace (x_{it}, y_{it}, D_{it})~|~ i\in\mathcal{I}, t\in\mathcal{T}\rbrace$. Let $\mathcal{S} = \lbrace 1, \ldots, n\rbrace$ denote the indices of the resulting training sample. Algorithm \ref{alg:causal_tree_recursive} provides the pseudocode used for fitting a causal tree.
\begin{algorithm}[h]
\caption{Recursive Causal Tree Algorithm \texttt{FitTree($\mathcal{S}$)}}
\label{alg:causal_tree_recursive}
\textbf{Input:} Set of sample point indices $\mathcal{S} = \lbrace 1, \ldots, n \rbrace$ \\
\textbf{Output:} \texttt{TreeNode root}
\begin{algorithmic}[1]
\IF{$n_{\text{treat}} < n_{\min}$ or $n_{\text{ctrl}} < n_{\min}$ or $\text{curr\textunderscore depth} \geq \text{max\textunderscore depth}$} \label{eq:if_condition}
\STATE return
\ELSE
\STATE Randomly create set of eligible features $\mathcal{J}$ \label{eq:bagging}
\STATE Choose best splitting feature $j\in \mathcal{J}$ and splitting threshold $\gamma$ \label{eq:choose_best_feature}
\STATE $\mathcal{S}_l = \lbrace i~|~x_{ij} \leq \gamma \rbrace$ \label{eq:left_data}
\STATE $\mathcal{S}_r = \lbrace i~|~x_{ij} > \gamma \rbrace$ \label{eq:right_data}
\STATE $\text{node\textunderscore left}$ = \texttt{FitTree($\mathcal{S}_l$)} \label{eq:recurse_left}
\STATE $\text{node\textunderscore right}$ = \texttt{FitTree($\mathcal{S}_r$)}\label{eq:recurse_right}
\ENDIF{}
\end{algorithmic}
\end{algorithm}

To avoid overfitting, Algorithm \ref{alg:causal_tree_recursive} stops growing the tree when the number of treated observations or control observations drops below a certain threshold $n_{\min}$ or if the current depth of the node exceeds the maximal depth $\text{max\textunderscore depth}$ (cf. line \ref{eq:if_condition}). If either of these three criteria is fulfilled, the current node is defined to be a leaf that predicts the CATE as the differences in means between treated and control observations:
\begin{equation}\label{eq:cate_computation}
\begin{split}
\hat{\beta}(\mathcal{S}) = \frac{1}{|\lbrace i|i\in\mathcal{S},D_i = 1 \rbrace|}\sum_{j\in \lbrace i|i\in\mathcal{S},D_i = 1 \rbrace}y_j \\
- \frac{1}{|\lbrace i|i\in\mathcal{S},D_i = 0 \rbrace|}\sum_{j\in \lbrace i|i\in\mathcal{S},D_i = 0 \rbrace}y_j.
\end{split}
\end{equation}
Otherwise, we use bagging \cite{Breiman:1984aa} to randomly generate an eligible subset of candidate splitting features $\mathcal{J}$ (line \ref{eq:bagging}). For each such feature, we iterate through various values to find the threshold that minimizes the weighted cost $J$ as follows:
\begin{align}\label{eq:weighted_cost}
J(\mathcal{S}) &= \frac{|\mathcal{S}_l| J(\mathcal{S}_l) + |\mathcal{S}_r| J(\mathcal{S}_r)}{|\mathcal{S}_l| + |\mathcal{S}_r|},
\end{align}
where the costs for the left and right subtree are defined as
\begin{subequations}
\begin{align}
J(\mathcal{S}_l) &= \text{mse}_l^0 + \text{mse}_l^1 - \alpha (\mu^0(\mathcal{S}_l) - \mu^1(\mathcal{S}_l))^2, \label{eq:cost_left} \\
J(\mathcal{S}_r) &= \text{mse}_r^0 + \text{mse}_r^1 - \alpha (\mu^0(\mathcal{S}_r) - \mu^1(\mathcal{S}_r))^2. \label{eq:cost_right}
\end{align}
\end{subequations}
In \eqref{eq:cost_left} and \eqref{eq:cost_right}, $\text{mse}_l^0$ and $\text{mse}_r^0$ denote the mean squared error of all control samples in the left and right subtree, respectively:
\begin{subequations}
\begin{align}
\text{mse}_l^0 &= \frac{\sum_{j\in \lbrace i|i\in\mathcal{S}_l,D_i = 0 \rbrace}\left( y_j - \mu^0(\mathcal{S}_l) \right)^2}{|\lbrace i|i\in\mathcal{S}_l,D_i = 0 \rbrace|}, \\
\text{mse}_r^0 &= \frac{\sum_{j\in \lbrace i|i\in\mathcal{S}_r,D_i = 0 \rbrace}\left( y_j - \mu^0(\mathcal{S}_r) \right)^2}{|\lbrace i|i\in\mathcal{S}_r,D_i = 0 \rbrace|}.
\end{align}
\end{subequations}
$\text{mse}_l^1$ and $\text{mse}_r^1$ are defined analogously. Next, the terms $\left( \mu^0(\mathcal{S}_l) - \mu^1(\mathcal{S}_l) \right)^2$ and $\left( \mu^0(\mathcal{S}_r) - \mu^1(\mathcal{S}_r) \right)^2$ denote the squared differences in means between control and treatment observations in the left and right subtree, respectively. Taken together, the cost function $J$ penalizes variations in the outcomes within the control or treatment group, but rewards separations between them. The scalar parameter $\alpha \in \mathbb{R}_+$ trades off these two contributions.

Finally, the sample $\mathcal{S}$ is split into $\mathcal{S}_l \cup \mathcal{S}_r$ according to the best splitting feature and threshold that minimizes the cost \eqref{eq:weighted_cost} and recursively calls \texttt{FitTree} on $\mathcal{S}_l$ and $\mathcal{S}_r$.

Taken together, Algorithm \ref{alg:causal_tree_recursive} partitions the feature space into a set of leaves $\Pi = \lbrace f_1, \ldots, f_{|\Pi|}\rbrace$, where each leaf $f\in\Pi$ has a prediction of the conditional CATE $\hat{\beta}(\mathcal{S}(f))$, where $\mathcal{S}(f)\subset \mathcal{S}$ denotes the collection of sample points in leaf $f$.

\subsection{Validation of Splitting Algorithm}
We now verify that algorithm \ref{alg:causal_tree_recursive} works correctly by simulating synthetic responses on a reduced data set. In doing so, we also find the optimal set of hyperparameters that minimizes the mean squared prediction error $\text{mse}$, which we define to be 
\begin{align}\label{eq:cate_mse}
\text{mse} = \frac{1}{|\mathcal{S}^{\text{val}}|} \sum_{f\in\Pi}\sum_{i\in f\subseteq\mathcal{S}^{\text{val}}} \left( \hat{\beta}(\mathcal{S}^{\text{tr}}(f)) - \beta_i \right)^2,
\end{align}
that is, the mean sum of squares between the actual treatment effect $\beta_i$ and the CATE predicted in the corresponding leaf $f$, i.e. $\hat{\beta}(\mathcal{S}^{\text{tr}}(f))$. By using a training sample $\mathcal{S}^{\text{tr}}$ that fits the causal tree $\Pi$ and predicts CATEs $\lbrace \hat{\beta}(\mathcal{S}^{\text{tr}}(f)) \rbrace_{f\in\Pi}$ as well as a separate validation set $\mathcal{S}^{\text{val}}$ to compute \eqref{eq:cate_mse}, we ensure the model does not overfit/underfit.

\begin{figure}[h]
\centering
\vspace*{-0.1cm}
\begin{tikzpicture}
\draw [ultra thick, draw=blue] (0,0) -- (4.975,0) -- (4.975,4) -- (0,4) -- (0,0);
\draw [ultra thick, draw=orange] (5.025,0) -- (7,0) -- (7,4) -- (5.025,4) -- (5.025,0);
\draw [ultra thick, draw=mygreen] (0,2.975) -- (7,2.975);
\draw [ultra thick, draw=red] (0,3.025) -- (7,3.025);

\node at (2.5,3) [circle, ultra thick, text=blue, draw=none] (tr) {{\LARGE $\mathbf{\mathcal{S}^{\text{tr}}}$}};
\node at (6,3) [circle, ultra thick, text=orange, draw=none] (val) {{\LARGE $\mathbf{\mathcal{S}^{\text{val}}}$}};

\fill[mygreen, opacity=0.4] (0,0) rectangle (7,3);
\fill[red, opacity=0.4] (0,3) rectangle (7,4);
\end{tikzpicture}
\vspace*{0.1cm}
\caption{Partitioning of data for cross-validation. First, apply \eqref{eq:synthetic_response} on 10\% of available data (red area) to create treatment data, remaining 90\% (green) is control data. Second, randomly choose 80\% of treatment and control data for training (blue rectangle, $\mathcal{S}^{\text{tr}}$), remaining 20\% is validation data (orange rectangle, $\mathcal{S}^{\text{val}}$).}
\label{fig:partitioning}
\end{figure}
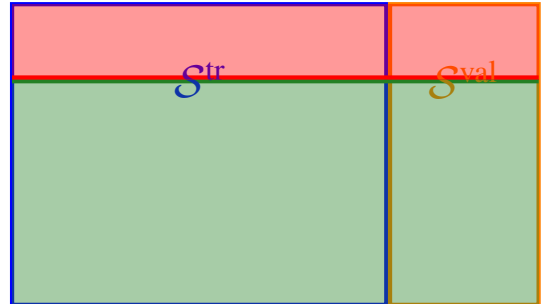

More precisely, we perform cross-validation on the hyperparameters $\alpha$ (cf. \eqref{eq:cost_left} and \eqref{eq:cost_right}) and $\text{max\textunderscore depth}$ to minimize \eqref{eq:cate_mse}, given a synthetic linear response which we construct as follows:
\begin{align}
y_{it} \leftarrow y_{it} &-(0.025\cdot T_{it} + \varepsilon_{T}) -(0.333\cdot \text{mean\textunderscore kwh}_i + \varepsilon_{m}) \nonumber\\
&-(\text{pugasheat}_i + \varepsilon_{pu}) - (\text{prenter}_i + \varepsilon_{pr}) \label{eq:synthetic_response}\\
&- (0.2\cdot \text{avgfamsize}_i + \varepsilon_{a}), \nonumber
\end{align}
where $\varepsilon_{T}, \varepsilon_{m}, \varepsilon_{pu}, \varepsilon_{pr}, \varepsilon_{a}$ are independent zero mean Gaussians with variance $\sigma^2 = 1/4$. The variables in \eqref{eq:synthetic_response} denote the temperature of user $i$ at time $t$, the mean consumption, the percentage of electricity consumed by gas heat, probability of renting a house, and the average family size. This subjectively chosen response is applied to a subset of users defined to be the treated group $\mathcal{T}$, whereas the remaining users become the control group $\mathcal{C}$. Next, we randomly draw 80\% of samples from both $\mathcal{T}$ and $\mathcal{C}$ to create the training set $\mathcal{S}^{\text{tr}}$, from which it follows that the remaining 20\% are the validation set $\mathcal{S}^{\text{val}}$, see Figure \ref{fig:partitioning} for an illustration.

\begin{figure}[hbtp]
\centering
\includegraphics[width=0.49\textwidth]{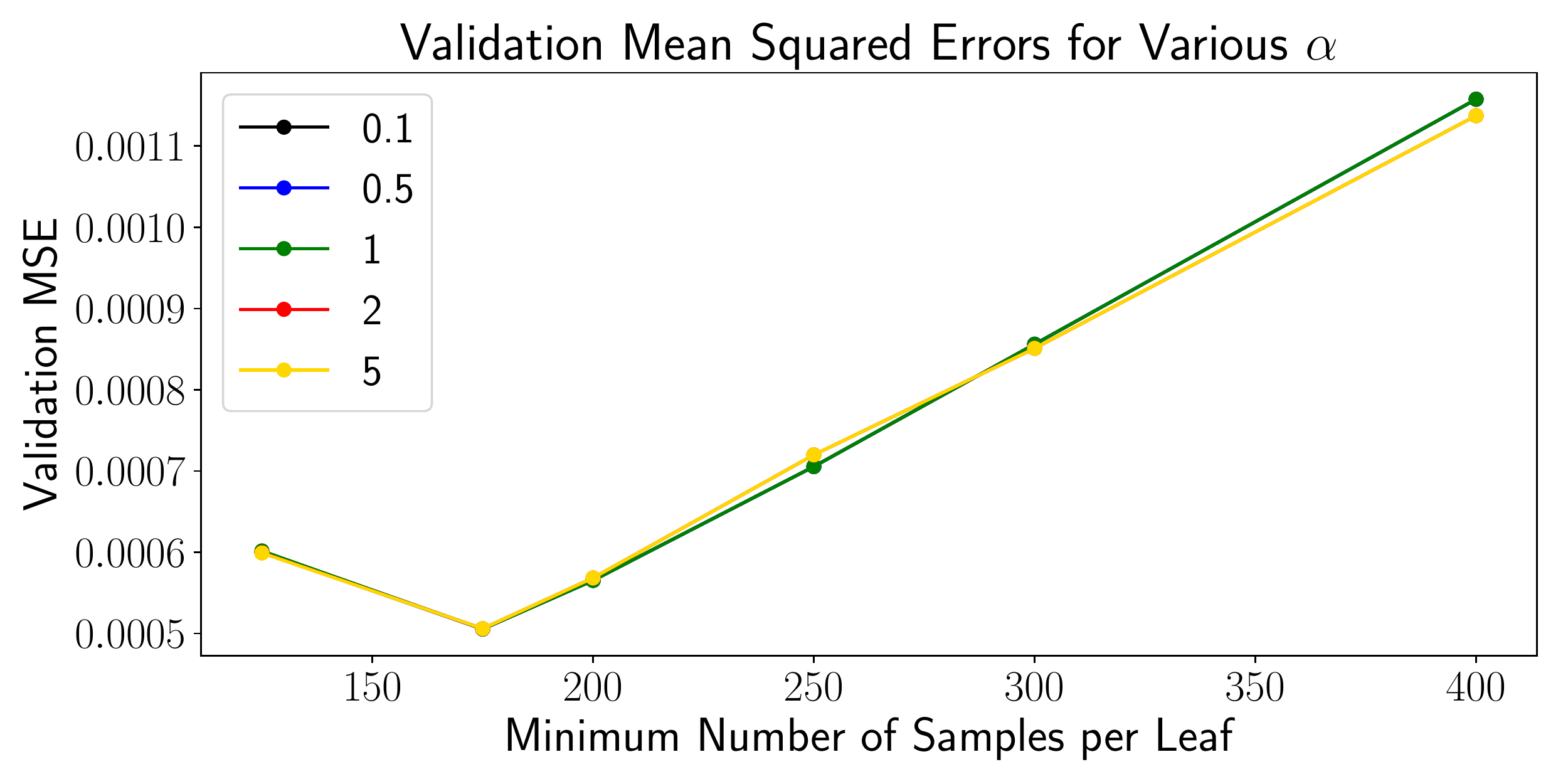}
\caption{Validation MSE \eqref{eq:cate_mse} for varying values of $\alpha$}
\label{fig:validation_mse}
\end{figure}

Figure \ref{fig:validation_mse} plots the validation MSE \eqref{eq:cate_mse} for various $\alpha$ and $n_{\min}$ with a maximal depth $\text{max\textunderscore depth} = 30$. Since the training and validation data set consist of separate data points, we observe that for $n_{\min} < 125$ some leaves will not be populated by samples in the validation set, resulting in an undefined MSE. As this phenomenon is a clear indication of overfitting, we define the MSE in this case to be infinity. Thus, in Figure \ref{fig:validation_mse}, all points to the left of a particular line have an infinite MSE. 

As the MSEs appear to be relatively insensitive to the choice of $\alpha$, particularly for $n_{\min}=175$ that minimizes the MSE, we will use $\alpha=1$ for all subsequent analyses of the entire data set.

\subsection{Results and Interpretation}
\begin{figure*}[hbtp]
\subfloat[]{\includegraphics[width=1.0\textwidth]{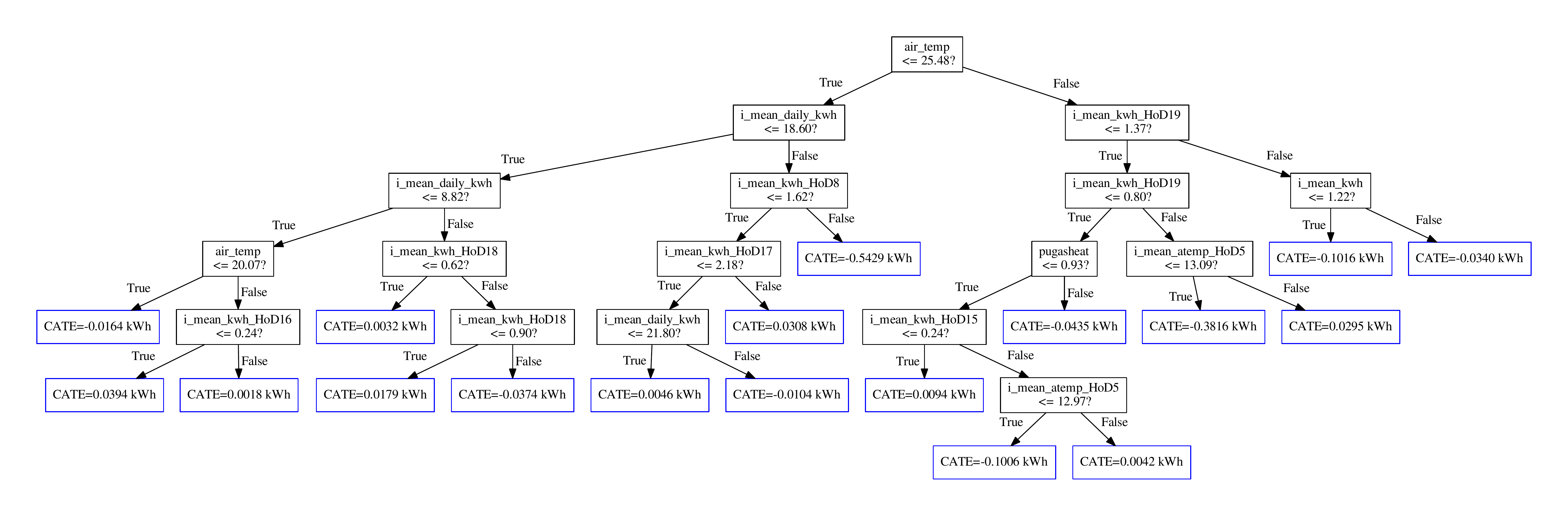}} \hfill \vspace*{-0.45cm}
\caption{Example for a causal tree on demeaned data set, maximal depth set to 6. Blue boxes (leaves) describe CATEs computed with \eqref{eq:cate_computation}.}
\label{fig:causal_tree_example}
\end{figure*}

We now perform the same analysis from the previous subsection on the full data set with synthetic responses across all covariates. Specifically, we divide the data into a training and validation set and choose the best set of hyperparameters that minimizes the MSE \eqref{eq:cate_mse}. As the training procedure presents no novel insights, we omit the learning curve for the MSE (cf. Figure \ref{fig:validation_mse}). Figure \ref{fig:causal_tree_example} plots an example for a causal tree that has been fitted with a subjectively chosen set of hyperparameters to fit within the page limit.
	
The optimal set of hyperparameters that achieves the smallest validation MSE \eqref{eq:cate_mse} was found to be $n_{\min} = 200$, $\text{max\textunderscore depth} = 15$. As in the previous subsection, the influence of the parameter $\alpha$ on the MSE seems to be negligible, motivating us to use $\alpha=1$ as in the validation case.

%
%

\section{K-Means Clustering for Heterogeneity}\label{sec:kmeans_clustering}
In the previous section we illustrated how a causal tree can detect heterogeneity in treatment effects. We now present a more ad-hoc approach which utilizes $k$-means clustering and compare its prediction accuracy to the mean squared error \eqref{eq:cate_mse} achieved by the causal tree. For a valid comparison, we use the same sample $\mathcal{S} = \mathcal{S}^{\text{tr}} \cup \mathcal{S}^{\text{val}}$ with the same synthetic responses \eqref{eq:synthetic_response} as in the causal tree case.

\begin{figure}[hbtp]
\centering
\includegraphics[width=0.49\textwidth]{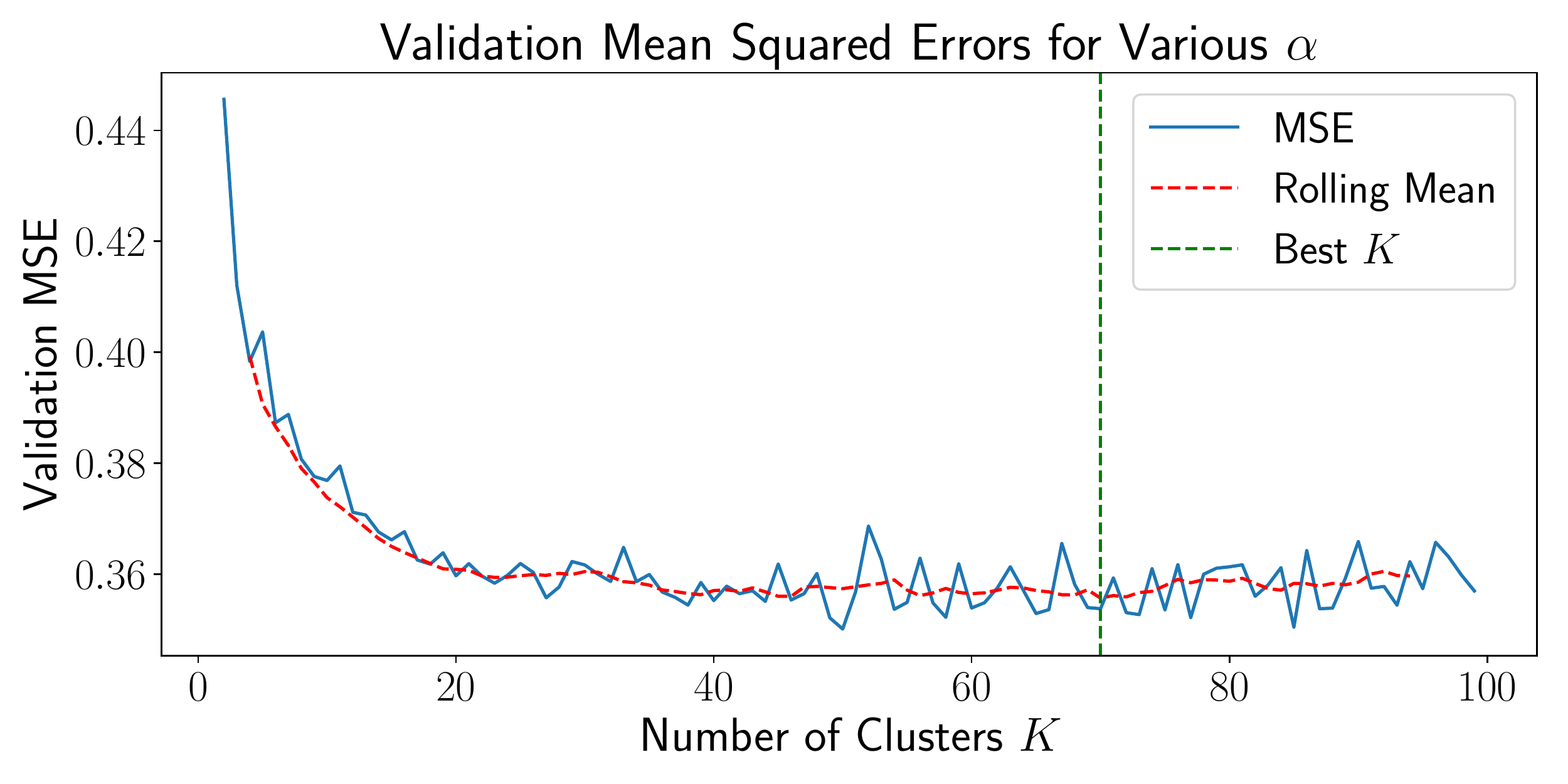}
\caption{Validation MSE \eqref{eq:cate_mse} for varying values of $K$}
\label{fig:kmeans_validation_mse}
\end{figure}

The well-known $k$-means clustering algorithm places $k$ centroids $c_1, \ldots, c_k$ into the $n$-dimensional feature space, where $n$ is the number of features of the training samples. The CATE belonging to cluster $c_i$ is then defined to be
\begin{equation}\label{eq:cate_computation_kmeans}
\begin{split}
\hat{\beta}(c_i) = \frac{1}{|\lbrace j\in\mathcal{S}^{\text{tr}}|c(j) = c_i, D_j = 1 \rbrace |}\sum_{k\in \lbrace j\in\mathcal{S}^{\text{tr}}|c(j) = c_i, D_j = 1\rbrace} \beta_k \\
- \frac{1}{|\lbrace j\in\mathcal{S}^{\text{tr}}|c(j) = c_i, D_j = 0 \rbrace |}\sum_{k\in \lbrace j\in\mathcal{S}^{\text{tr}}|c(j) = c_i, D_j = 0\rbrace} \beta_k
\end{split}
\end{equation}
which is similar in nature to \eqref{eq:cate_computation}. In \eqref{eq:cate_computation_kmeans}, $c(j)$ denotes the index of the centroid with the smallest Euclidian distance to training sample $j\in\mathcal{S}^{\text{tr}}$. Hence the $k$-means algorithm generates $k$ distinct CATEs, namely one for each cluster. Finally, the MSE is determined as the mean of squared differences between predicted and actual treatment effect:
\begin{align}
\text{mse} = \frac{1}{|\mathcal{S}_{\text{val}}|} \sum_{i\in\mathcal{S}^{\text{val}}} \left( \hat{\beta}(c(i)) - \beta_i \right)^2.
\end{align}
To find the optimal value of $k$, we perform cross-validation on a range of possible $k$, see Figure \ref{fig:kmeans_validation_mse}. 
In addition to the MSEs, Figure \ref{fig:kmeans_validation_mse} also plots their rolling mean with a window size of $8$, as the MSE oscillates around its trend. Using this technique, we determine the optimal $k$ to be 70.

Comparing Figures \ref{fig:validation_mse} and \ref{fig:kmeans_validation_mse}, we immediately see that the validation MSE of the causal tree is two orders of magnitude smaller than in the $k$-means clustering case, which is consistent with our expectations, as the $k$-means clustering algorithm fails to take into account the labels of the data points and operates on a much more naive level than the causal tree.

%
%

\section{Discussion and Conclusion}\label{sec:conclusion}
We analyzed Residential Demand Response as a human-in-the-loop cyber-physical system that incentivizes users to curtail electricity consumption during designated peak hours. Utilizing data collected from a Randomized Controlled Trial funded by the California Energy Commission and conducted by a DR provider in the San Francisco Bay Area, we estimated the causal effect of hour-ahead price interventions on electricity reduction. 

We developed a two-step \textit{non-experimental} estimation framework and employed off-the-shelf regression models to learn a consumption model on non-DR periods to predict counterfactuals during DR hours of interest. The estimated ATE is $-0.10$ kWh (11\%) per Demand Response event and household. Further, we observe a weak positive correlation between the incentive level and the estimated reductions, suggesting that users are only weakly elastic in response to incentives.

Moreover, we discovered notable heterogeneity of users in time and by automation status, since the largest reductions were observed in summer months as well as among users with at least one connected smart home automation device. Further, the ambient air temperature was found to positively correlate with the amount of reductions, suggesting that air conditioning units are a major contributor to reductions.

Further, we used causal decision trees and $k$-means clustering to detect subpopulations of ``similar'' treatment effects. While the $k$-means algorithm is a naive algorithm, fitting a causal decision trees requires more complex computations, resulting in much higher accuracies. In our previous work \cite{Zhou:2018ab}, we made use of a very simplistic targeting algorithm, which assigns large incentives to small reducers and vice versa in order to reduce the cost of the DR provider. Future work will investigate to what degree cost efficiency can be improved by utilizing more fine-grained targeting approaches enabled by causal decision trees.




\bibliographystyle{IEEEtran}
\bibliography{bibliography}

\begin{IEEEbiography}[{\includegraphics[width=1in,height=1.25in,clip,keepaspectratio]{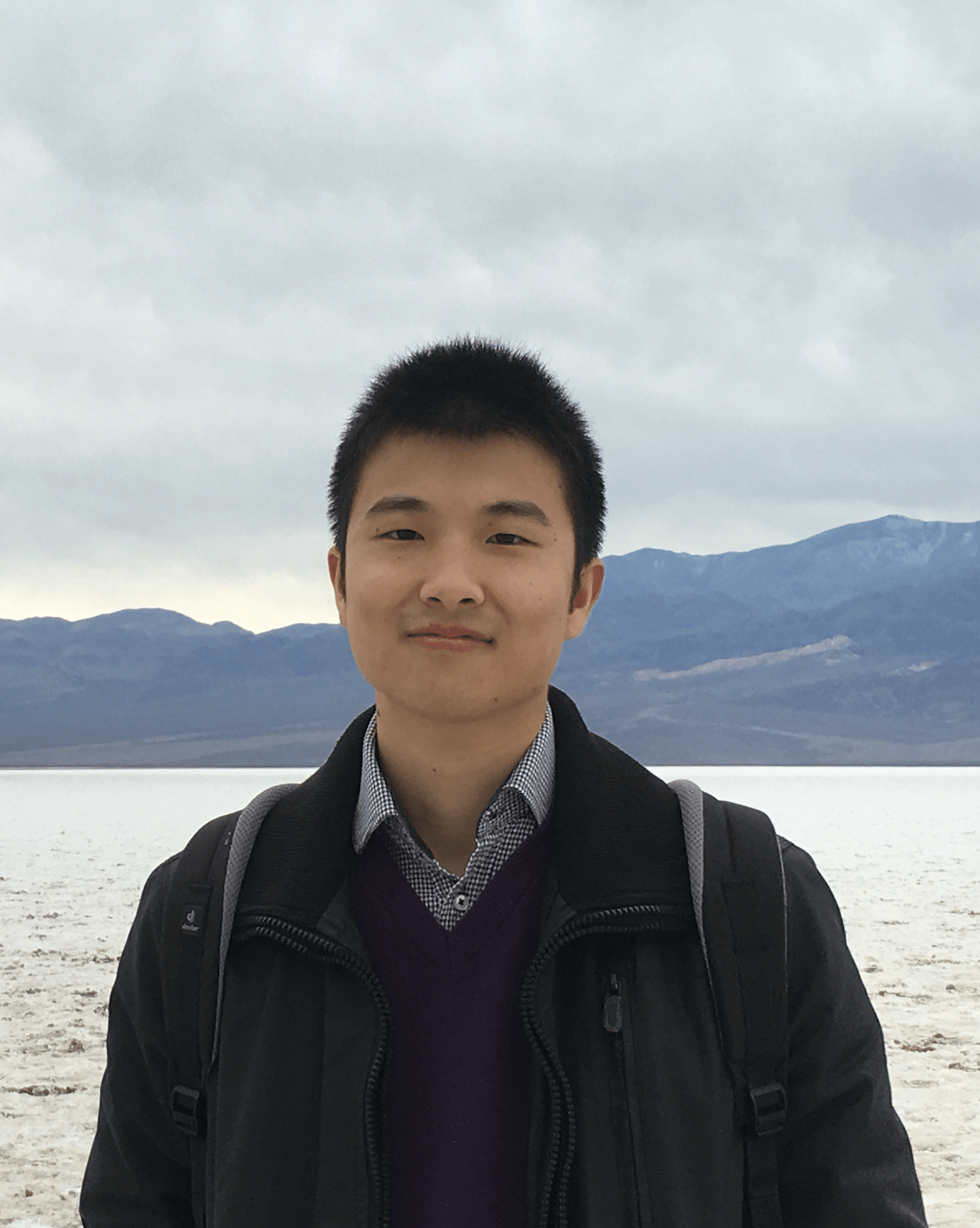}}]{Datong Paul Zhou}
received a B.S. in Mechanical Engineering from TU Munich, Germany, in 2014. He received M.S. degrees in Electrical Engineering and Computer Science and Mechanical Engineering from the University of California, Berkeley, in 2016 and 2017, respectively, as well as an M.A. in Mathematics in 2018. He is currently working towards his Ph.D. in Mechanical Engineering with a focus on machine learning, game theory, and control theory applied to electricity markets. 
\end{IEEEbiography}

\begin{IEEEbiography}[{\includegraphics[width=1in,height=1.25in,clip,keepaspectratio]{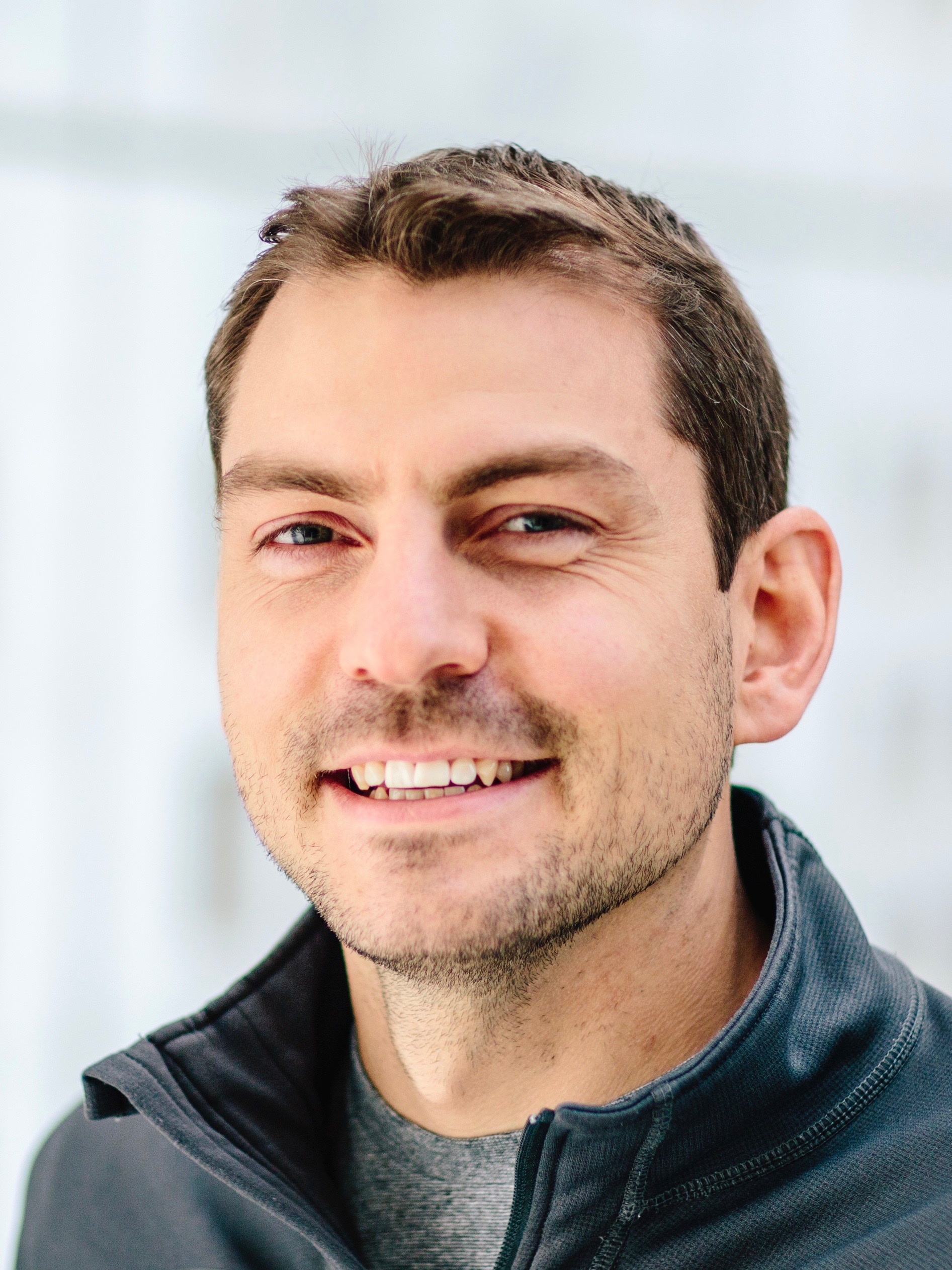}}]{Maximilian Balandat}
received a Dipl.-Ing. in Electrical Engineering from TU Darmstadt, Germany, in 2010, and a M.A. in Mathematics and a Ph.D. in Electrical Engineering from UC Berkeley in 2016. His research interests include Control Theory, Machine Learning, Causal Inference, Economic Incentives, and Bayesian Optimization. He is currently a Research Scientist at Facebook.
\end{IEEEbiography}

\begin{IEEEbiography}[{\includegraphics[width=1in,height=1.25in,clip,keepaspectratio]{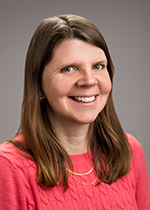}}]{Claire Jennifer Tomlin}
is the Charles A. Desoer Professor of Engineering in EECS at Berkeley. She was an Assistant, Associate, and Full Professor in Aeronautics and Astronautics at Stanford from 1998 to 2007, and in 2005 joined Berkeley. Claire works in the area of control theory and hybrid systems, with applications to air traffic management, UAV systems, energy, robotics, and systems biology. She is a MacArthur Foundation Fellow (2006), an IEEE Fellow (2010), and in 2017 was awarded the IEEE Transportation Technologies Award.
\end{IEEEbiography}

%

\end{document}